\documentclass[aps,pra,twocolumn,showpacs,preprintnumbers,letterpaper]{revtex4}
\usepackage{graphicx}
\usepackage{amsmath}
\usepackage{epsfig}
\usepackage{amssymb}

\begin{document}

\title{Energetics and Structural Properties
of Trapped Two-Component Fermi Gases}

\author{J. von Stecher}
\affiliation{Department of Physics and JILA, University of Colorado,
Boulder, CO 80309-0440}
\author{Chris H. Greene}
\affiliation{Department of Physics and JILA, University of Colorado,
Boulder, CO 80309-0440}
\author{D. Blume}
\affiliation{JILA, University of Colorado,
Boulder, CO 80309-0440}
\affiliation{Department of Physics and Astronomy,
Washington State University,
  Pullman, Washington 99164-2814}

\date{\today}

\begin{abstract}
Using two different numerical methods, we study the behavior of
two-component Fermi gases interacting through short-range $s$-wave
interactions in a harmonic trap. A correlated Gaussian basis-set
expansion technique is used to determine the energies and structural
properties, i.e., the radial one-body densities and pair
distribution functions, for small systems with either even or odd
$N$, as functions of the $s$-wave scattering length and the mass
ratio $\kappa$ of the two species. Particular emphasis is put on a
discussion of the angular momentum of the system in the BEC-BCS
crossover regime. At unitarity, the excitation spectrum of the
four-particle system with total angular momentum $L=0$ is calculated
as a function of the mass ratio $\kappa$. The results are analyzed
from a hyperspherical perspective, which offers new insights into
the problem. Additionally, fixed-node diffusion Monte Carlo
calculations are performed for equal-mass Fermi gases with up to
$N=30$ atoms. We focus on the odd-even oscillations of the ground
state energy of the equal-mass unitary system having up to $N=30$
particles, which are related to the excitation gap of the system.
Furthermore, we present a detailed analysis of the structural
properties of these systems.
\end{abstract}

\maketitle

\section{Introduction}

Pure Fermi systems with essentially any interaction strength can be
realized experimentally with ultracold atomic gases. In most
experiments to date, large samples of atomic Li or K are trapped
optically in two different hyperfine states, in the following simply
referred to as ``spin-up'' and ``spin-down'' states. By tuning an
external magnetic field in the vicinity of a Fano-Feshbach
resonance~\cite{stwa76a,ties93,inou98,corn00}, the interspecies
$s$-wave scattering length can be varied from non-interacting to
infinitely strongly-interacting (either attractive or repulsive).
This tunability is unique to atomic systems, and it has enabled for
the first time quantitative experimental studies of the crossover
from the molecular BEC-regime to the atomic
BCS-regime~\cite{grei03,zwie03,bour03,stre03,kina04,stew06}. Since
the systems studied experimentally are in general large, many
observations have been explained quite successfully by applying
theoretical treatments based on the local density
approximation (LDA); see, e.g., Ref.~\cite{gior07} and references
therein. The LDA uses the equation of state of the homogeneous
system as input, and, in general, accurately describes the
properties of the system near the trap center, where the density
changes slowly.  However, it fails to accurately describe the
properties of the system near the edge of the cloud, where the
density varies more rapidly.

In a different set of experiments, atomic Fermi gases are loaded
into an optical lattice with variable barrier
height~\cite{kohl05,mori05,rom06}. In the regime where the tunneling
of atoms between neighboring lattice sites can be neglected, each
lattice site provides an approximately harmonic confining potential
for the atoms at that site. Through the application of a so-called
``purification scheme''~\cite{thal06}, experimentalists are now able
to realize systems with a deterministic number of atoms per site. So
far, optical lattices have been prepared with one or zero atoms per
site, with two or zero atoms per site, and with three or zero atoms
per site. Optical-lattice experiments thus allow for the
simultaneous preparation of multiple copies of identical
few-particle systems. We anticipate that these experiments will be
extended to larger atom samples in the future, thereby opening the
possibility to study systematically how the properties of the system
change as functions of the number of atoms. Transitions from few- to
many-body systems have, e.~g., been studied experimentally in metal
and rare gas clusters~\cite{heer93,toen04}, and it is exciting
that the experimental study of this transition in dilute gaseous
systems is within reach.  A mature body of theoretical work has also
investigated the manner in which bulk electronic, magnetic and
superfluid properties can be understood by studying small or
modest-size clusters~\cite{LeeCallawayDhar1984,sind89}.

This paper presents theoretical results for trapped two-component
Fermi gases with up to $N=30$ fermions, which shed light on the few-
to many-body transition from a microscopic or few-body point of
view. To solve the many-body Schr\"odinger equation we use two
different numerical methods, a correlated Gaussian (CG) basis set
expansion approach and a fixed-node diffusion Monte Carlo (FN-DMC)
approach. The CG approach allows for the determination of the entire
energy spectrum and eigenstates with controlled accuracy (i.e., no
approximations are employed and the convergence can be
systematically improved). If we demand an accuracy of the order of
2\% or better, our current CG implementation limits us to treating
systems with up to $N=6$ atoms (and to the lowest 10 or twenty eigen
states). To the best of our knowledge, no other such calculations
exist for dilute fermionic few-body systems ($N =4-6$) with
short-range interactions. The FN-DMC method, in contrast, can be
applied to larger systems but its accuracy crucially depends on the
quality of the many-body nodal surface, which is in general unknown.
Moreover, the FN-DMC approach as implemented here treats only ground
state properties for the chosen symmetry. Careful comparisons of the
ground state energy and structural properties calculated by the
FN-DMC and CG approach for different interaction strengths validate
the construction of the nodal surfaces employed for $N \le6$. We
expect, and provide some evidence, that our nodal surfaces
constructed to describe the energetically lowest-lying gas-like
state of larger $N$ are also quite accurate.

Specifically, we calculate the energy of the energetically
lowest-lying gas-like state of trapped two-species Fermi gases as a
function of the number of particles $N$, the $s$-wave scattering
length $a_s$ and the mass ratio $\kappa$. Our ground state energies
for even and odd $N$ can be readily combined to determine the
excitation gap, which is related to pairing physics. For small
systems, we additionally determine and discuss the excitation
spectrum. Furthermore, we present pair correlation functions, which
provide further insights into the pair formation process, and radial
density profiles for the ground state. Finally, we elaborate on the
interpretation of the behaviors within a framework that uses
hyperspherical coordinates. This connection has been summarized in
an earlier paper~\cite{blumePRL07}. Here, we present additional
results and discuss in more detail how the even-odd oscillations
emerge in the hyperspherical framework. Our analysis provides an
alternative means, complementary to conventional many-body theory,
for understanding the excitation gap at unitarity.

The remainder of this paper is organized as follows.
Section~\ref{sec_theory} introduces the Hamiltonian of the
system under study, reviews the definitions of the
normalized energy crossover curve and the excitation gap,
and summarizes some peculiar
properties of the unitary gas using hyperspherical coordinates.
Section~\ref{sec_method} summarizes the CG and FN-DMC approaches,
and provides some implementation details specific to the problem at hand.
Section~\ref{sec_results} presents our results for the ground state
energies,
the excitation spectrum and structural properties.
Finally, Sec.~\ref{sec_conclusion} concludes.

\section{Theoretical background}
\label{sec_theory}
\subsection{Hamiltonian}
\label{sec_ham}
The main objective of this article is to obtain and interpret solutions to the
many-body time-independent
Schr\"odinger equation for a trapped
two-component Fermi gas with short-range interactions.
The model Hamiltonian for $N_1$
fermions of mass $m_1$ and $N_2$ fermions of mass $m_2$ reads
\begin{eqnarray}
\label{eq_ham} H = \sum_{i=1}^{N_1} \left(\frac{-\hbar^2}{2m_1}
\nabla_i^2 + \frac{1}{2} m_1 \omega^2 \vec{r}_i^2 \right) +
\nonumber \\
\sum_{i'=1}^{N_2} \left( \frac{-\hbar^2}{2m_2} \nabla_{i'}^2 +
\frac{1}{2} m_2 \omega^2 \vec{r}_{i'}^2 \right) +\sum_{i=1}^{N_1}
\sum_{i'=1}^{N_2} V_0(r_{ii'}).
\end{eqnarray}
Here, $\vec{r}_i$ and $\vec{r}_{i'}$ denote the position vector of
the $i$th mass $m_1$ fermion and the $i'$th mass $m_2$ fermion,
respectively. Both atom species experience a trapping potential
characterized by the same angular frequency $\omega$. For equal
masses, this is indeed the case in ongoing experiments. For unequal
masses, however, the two atomic species typically experience
different trapping frequencies. Our restriction to equal trapping
frequencies reduces the parameter space which otherwise would be
impractical to explore numerically. Furthermore, our CG calculations
simplify for equal trapping frequencies because the center-of-mass
and relative motions decouple in this case. The studies presented
here for unequal masses but equal frequencies complement our earlier
study~\cite{vonstechtbp}, which treats two-component Fermi gases
with unequal masses that experience trapping frequencies $\omega_1$
and $\omega_2$ adjusted so that $m_1 \omega_1=m_2 \omega_2$. In
Eq.~(\ref{eq_ham}), $V_0$ is a short-range two-body potential
between each pair of mass $m_1$ and mass $m_2$ atoms. We
characterize the strength of $V_0$ by the $s$-wave scattering length
$a_s$, which can be varied experimentally through the application of
an external magnetic field in the vicinity of a Fano-Feshbach
resonance. Here, we model this situation by changing the depth of
$V_0$; our results should be applicable to systems with a broad
$s$-wave Fano-Feshbach resonance and vanishingly small $p$-wave
interactions.

The present study considers two-component Fermi gases with either
even or odd $N$, where $N=N_1+N_2$.Because odd-even oscillations
serve as one major subject of this study, we set $N_1=N_2$ for even
$N$, and $N_1=N_2\pm1$ for odd $N$. In addition to the scattering
length $a_s$, we vary the mass ratio $\kappa$,
\begin{eqnarray}
\kappa = m_1/m_2.
\end{eqnarray}
Throughout, we take $m_1 \ge m_2$ so that $\kappa \ge 1$. In most
cases, we measure lengths in units of the oscillator length
$a_{ho}$, $a_{ho}=\sqrt{\hbar/(2\mu\omega)}$, which  is defined in
terms of the reduced mass $\mu$, $\mu=m_1 m_2/(m_1+m_2)$.

It has been shown previously~\cite{petr04,petr05,petr05a,blumePRL07}
that small equal-mass two-component Fermi gases, which interact
through short-range two-body potentials with infinitely large $a_s$
that support no $s$-wave bound state, support no tightly-bound
many-body states with negative energy. For unequal mass systems the
situation is
different~\cite{efim70,efim73,petr05,kartavtsev2007let}. Trimers
consisting of two heavy particles and one light particle that
interact through short-range potentials support tightly-bound states
with negative energy if the mass ratio and the scattering length are
sufficiently large. Reference~\cite{vonstechtbp} discussed the role
of non-universal trimer states for unequal-mass systems in some
detail, and we return to this discussion in
Sec.~\ref{sec_energyresult}. Throughout this work, we restrict our
analysis to gas-like states, consisting of atomic fermions,
molecular bosons or both.

To solve the Schr\"odinger equation for eigenstates of $H$, we use
two different numerical methods: a correlated Gaussian (CG) basis
set expansion technique and a fixed-node diffusion Monte Carlo
(FN-DMC) technique. For numerical convenience, we utilize different
short-range potentials $V_0$ in our CG and FN-DMC calculations. We
adopt a purely attractive Gaussian interaction potential defined as
\begin{eqnarray}
V_0(r)=-d
\exp \left(- \frac{r^2}{2R_0^2} \right)
\end{eqnarray}
in the CG calculations, and a
square well interaction potential defined as
\begin{eqnarray}
V_0(r) = \left\{ \begin{array}{cl}
-d & \mbox{for } r < R_0 \\
0 & \mbox{for } r > R_0
\end{array} \right.
\end{eqnarray}
in the FN-DMC calculations. For a fixed range $R_0$, the potential
depth $d$ is adjusted so that the $s$-wave scattering length $a_s$
takes the desired value. The range $R_0$ is selected so that $R_0\ll
a_{ho}$. The premise is that the properties of two-component Fermi
gases with short-range interactions (or at least the universal state
properties) are determined by the $s$-wave scattering length $a_s$
alone, and independent of the details of the underlying two-body
potential if the range $R_0$ is chosen sufficiently small. Ideally,
we would consider the limit $R_0 =0$. This is, however, impossible
within the numerical frameworks employed. Thus, we perform
calculations for different finite $R_0$, which allows us to
approximately extrapolate to the $R_0 =0$ limit and to estimate the
dependence of our results on $R_0$, i.e., to estimate the scale of
the finite-range effects.

\subsection{Energy crossover curve and excitation gap}
\label{sec_energy} The energetically lowest-lying gas-like states of
two-component Fermi gases with short-range interactions determine
the normalized energy crossover curve $\Lambda_N^{(\kappa)}$ and the
excitation gap $\Delta(N)$. To simplify the notation, the
energetically lowest-lying gas-like state is referred to as the
ground state in this section. The BCS and BEC limits of the
crossover can be treated perturbatively. For small $|a_s|$ and $a_s
< 0$, the system behaves like a weakly-interacting atomic Fermi gas
whose leading order properties beyond the non-interacting degenerate
Fermi gas are determined by $a_s$. For attractive two-body
potentials that generate small $a_s$ and $a_s>0$, in contrast, the
system behaves like a weakly-interacting molecular Bose gas whose
properties are to leading order determined by $a_{dd}$, where
$a_{dd}$ denotes the dimer-dimer scattering length.  (One can also
have small, positive $a_s$ with purely repulsive two-body potentials
that have no bound molecular states, but these systems behave quite
differently and will not be considered in this paper.) In the
strongly-interacting regime (large $|a_s|$), perturbation theory
cannot be applied and it is not clear {\em{a priori}} whether the
system behaves more like an atomic gas or a molecular gas, or like
neither of the two.

The definition of the normalized energy crossover curve
$\Lambda_{N_1,N_2}^{(\kappa)}$ introduced in
Refs.~\cite{vonstechtbp,jauregui07} for even $N$ can be extended to
odd $N$,
\begin{eqnarray}
\label{eq_cross} \Lambda_{N_1,N_2}^{(\kappa)} =
\frac{E(N_1,N_2)-N_{d}E(1,1)-3N_f/2\hbar\omega}
{E_{NI} - \frac{3}{2} N \hbar \omega}.
\end{eqnarray}
Here,
$E(N_1,N_2)$ denotes the ground state energy of the trapped
two-component gas consisting of $N_1$ fermions with mass $m_1$
and $N_2$ fermions with mass $m_2$.
In Eq.~(\ref{eq_cross}),
$N_d$ is defined by
\begin{eqnarray}
N_d=\min\{N_1,N_2\},
\end{eqnarray}
and corresponds to the number of dimers formed on the BEC side,
i.e., in the regime where $a_s$ is small and positive.
$N_f$ is defined by
\begin{eqnarray}
N_f=|N_1-N_2|;
\end{eqnarray}
it represents the number of unpaired atoms on the BEC side, and
takes the value 0 for even $N$ and 1 for odd $N$.

In Eq.~(\ref{eq_cross}), $E_{NI}$ denotes the ground state energy of
the non-interacting two-component Fermi gas consisting of $N$ atoms,
where--- as before--- $N=N_1+N_2$. The $E_{NI}$ can be evaluated as
the sum of the noninteracting energies of polarized Fermi gases
$E^p_{NI}$ with $N_1$ and $N_2$ particles,
$E_{NI}(N)=E^p_{NI}(N_1)+E^p_{NI}(N_2)$. Following
Ref.~\cite{Schneider98}, the $E^p_{NI}(N_i)$ can be written in terms
of the shell number $n_s$, the energy of the closed shell subsystem
$E^{cs}_{NI}(n_s)$, and the corresponding magic number $N^{cs}$,
\begin{equation}
E^p_{NI}(N_i)=E^{cs}_{NI}(n_s)+\left(\frac{3}{2}+n_s\right)(N-N^{cs})\hbar\omega.
\end{equation}
The shell number $n_s$ represents the number of closed shells and is
given by
\begin{equation}
n_s=\mbox{Int}\left[\frac{1}{g(N_i)}+\frac{g(N_i)}{3}-1\right],
\end{equation}
where
\begin{equation}
g(N_i)=\sqrt[3]{3\left(27 N_i-\sqrt{3(243 N_i^2-1)}\right)}
\end{equation}
and Int[$x$] is the integer part of $x$. Finally, the energy of the
closed shell subsystem $E^{cs}_{NI}(n_s)$ and the corresponding
magic number $N^{cs}$ are
\begin{gather}
N^{cs}=\frac{n_s(n_s+1)(n_s+2)}{6}\,\,\,\,\,\mbox{and}\\
\frac{E^{cs}_{NI}(n_s)}{\hbar\omega}=\frac{(n_s-1)n_s(n_s+1)(n_s+2)}{8}+\frac{3N^{cs}}{2}.
\end{gather}

On the positive $a_s$ side where a high-lying two-body bound state
exists, a significant fraction of the ground state energy of the $N$
fermion system is determined by the binding energy of the trapped
dimer, which depends on $R_0$. To reduce the dependence of
$\Lambda_{N_1,N_2}^{(\kappa)}$ on the range $R_0$, the energy
$E(1,1)$ of $N_d$ trapped dimer pairs is subtracted in
Eq.~(\ref{eq_cross}). Thus, $\Lambda_{N_1,N_2}^{(\kappa)}$ depends
to a good approximation only on $a_s$, $\kappa$, $N_1$ and $N_2$,
and not on the details of the underlying two-body potential (see
also Sec.~\ref{sec_energyresult}). By construction,
$\Lambda_{N_1,N_2}^{(\kappa)}$ changes from one on the
weakly-interacting BCS side (small $|a_s|$ and $a_s<0$) to zero on
the weakly-interacting molecular BEC side (small, positive $a_s$).

The weakly-interacting regimes, where $|a_s| \ll a_{ho}$, can be
treated perturbatively assuming zero-range interactions, i.e., a
Fermi pseudopotential~\cite{ferm34}. For small $|a_s|$ and $a_s <0$,
the energy within first order perturbation theory becomes
\begin{eqnarray}
\label{eq_expbcs} E \approx E_{NI}  + \hbar \omega \,
C_{N_1,N_2}^{\kappa} \, \frac{a_s}{a_{ho}},
\end{eqnarray}
where $C_{N_1,N_2}^{\kappa}$ is a dimensionless quantity. In
general, the evaluation of $C_{N_1,N_2}^{\kappa}$ is a bit
cumbersome since there is no unique ground state and degenerate
perturbation theory must be applied. When both $N_1$ and $N_2$
correspond to closed shells, then $C_{N_1,N_2}^{\kappa}$ can be
calculated straightforwardly analytically~\cite{vonstechtbp},
\begin{equation}
  \label{Cnkappa}
 C_{N_1,N_2}^{\kappa} =4 \pi a_{ho}^3 \int \rho_1^{NI}(\vec{r}) \rho_2^{NI}(\vec{r})  d\vec{r}.
\end{equation}
Here, $\rho_i^{NI}(\vec{r})$ is the density of a one-component
non-interacting gas with $N_i$ fermions of mass $m_i$, normalized so
that $\int \rho_i^{NI}(\vec{r}) d \vec{r}=N_i$. Alternatively, one
can approximate the $\rho_i^{NI}$ by the Thomas-Fermi density
profiles. This approximation should be quite accurate in the large
$N$ limit.

\begin{figure}[h]
\includegraphics[scale=0.5]{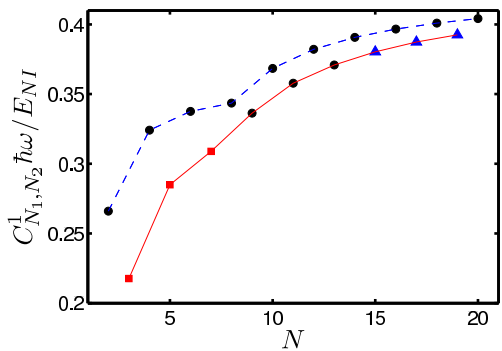}
\caption{(Color Online) $C_{N_1,N_2}^{1}$ coefficients divided by
$E_{NI}$ as a function of $N$. Circles correspond to $L=0$ ground
state, squares to $L=1$ ground state and triangles to $L=2$ ground
state. A solid line connects the odd-$N$ values while a dashed line
connects the even-$N$ values. } \label{Cnn}
\end{figure}

To obtain the $C_{N_1,N_2}^{\kappa}$ for open-shell systems, we
apply first-order degenerate perturbation theory. This calculation
additionally allows us to obtain the angular momentum quantum number
$L$ of the ground state. Figure~\ref{Cnn} and Table~\ref{table1}
present the results for $N\le20$ and $\kappa=1$. The coefficients
$C_{N_1,N_2}^{1}$ increase monotonically with increasing $N$ and
show a slight odd-even staggering. In general, the coefficients
$C_{N_1,N_2}^{1}$ for even $N$ are comparatively higher than those
for odd $N$, implying a smaller energy for even $N$ than for odd $N$
and suggesting that, even in the perturbative regime, the odd-even
oscillations are already present. We note that the $C_{N_1,N_2}^1$
coefficients for even $N$ shown in Fig.~\ref{Cnn} clearly reflect
the shell-closure at $N=8$.

\begin{table}
\caption{\label{table1} Angular momentum $L$ and coefficient
$C_{N_1,N_2}^{1}$ for the ground state of equal-mass two-component
Fermi gases in the weakly-attractive regime. Here, we consider
$N_2=N_1$ for even $N$ and $N_1=N_2+1$ for odd $N$.}
\begin{ruledtabular}
\begin{tabular}{c|ll||c|ll}
  $N$ & $L$ & $C_{N_1,N_2}^{1}$ & $N$ & $L$ & $C_{N_1,N_2}^{1}$\\
\hline
2& 0 &  $\frac{2}{\sqrt{2\pi}}$   & 12   & 0 &12.2274  \\
3& 1 &  $\frac{3}{\sqrt{2\pi}}$    & 13   & 0 &13.1651  \\
4& 0&  $\frac{13}{2\sqrt{2\pi}}$    & 14   & 0 &15.2382  \\
5& 1 & $\frac{15}{2\sqrt{2\pi}}$     & 15   & 2 & 16.1642\\
6& 0 &  $\frac{11}{\sqrt{2\pi}}$  & 16   & 0 &18.2445 \\
7& 1 &  $\frac{12}{\sqrt{2\pi}}$    & 17   & 2 &19.1735 \\
8&  0 &  $\frac{31}{2\sqrt{2\pi}}$  & 18   & 0 & 21.2476\\
9&  0 & $\frac{145}{8\sqrt{2\pi}}$  & 19   & 2 & $\frac{1779}{32\sqrt{2\pi}}$ \\
10& 0 &  9.21052  &  20  & 0 & $\frac{1945}{32\sqrt{2\pi}}$ \\
11& 0 &  10.1980 &       & & \\
\end{tabular}
\end{ruledtabular}
\label{EnU4}
\end{table}

In the weakly-interacting molecular BEC regime, the two-component
Fermi system should behave like a system that consists of $N_d$
bosonic molecules and $N_f=0$ or $1$ fermions. In first order
perturbation theory the ground state energy of such a system is
given by
\begin{eqnarray}
\label{eq_expbec} E \approx N_d E(1,1) +
\hbar\omega \frac{3
N_f}{2}
+
 \hbar \omega \, \frac{N_d(N_d-1)}{2}
\sqrt{\frac{2}{\pi}} \, \frac{a_{dd}}{a_{ho}^{(dd)}}\nonumber\\
 +\hbar \omega \,
N_d N_f \sqrt{\frac{2}{\pi}} \, \frac{a_{ad}}{a_{ho}^{(ad)}}.
\end{eqnarray}
Here, $a_{dd}$ and $a_{ad}$ denote the dimer-dimer and
atom-dimer scattering lengths, respectively.
The oscillator lengths $a_{ho}^{(dd)}$ and $a_{ho}^{(ad)}$
for the dimer-dimer and atom-dimer systems,
$a_{ho}^{(dd)}=\sqrt{\hbar/(2\mu_{dd}\omega)}$ and
$a_{ho}^{(ad)}=\sqrt{\hbar/(2\mu_{ad}\omega)}$,
are defined in terms of the
reduced mass $\mu_{dd}$ of the dimer-dimer system and
the reduced mass $\mu_{ad}$ of the atom-dimer system, respectively.

The limiting behaviors of the BEC-BCS crossover curve can be used to
guide the construction of the many-body nodal surface, which is a
crucial ingredient for our FN-DMC calculations (see
Sec.~\ref{sec_fndmc}). In the weakly-interacting molecular BEC
regime, even $N$ systems consist of $N/2$ dimers. Each molecule is
expected to be in its rotational ground state, leading to a
many-body wave function with total angular momentum $L=0$. For odd
$N$ systems, the extra fermion is expected to occupy the lowest
$s$-wave orbital, leading, as in the even $N$ case, to a many-body
wave function with $L=0$. Thus, the angular momentum of even $N$
systems is expected to be the same along the crossover while that of
odd $N$ systems is expected to change (for $N=3$, this has been
pointed out recently by two independent
groups~\cite{kestnerPra07,StetcuCM07}). This symmetry change
introduces a kink in the normalized energy curve
$\Lambda_{N_1,N_2}^{(\kappa)}$ for odd $N$ and in the excitation gap
$\Delta(N)$ (see below) at the scattering length where the symmetry
change or inversion occurs.

In addition to the energy crossover curve, we calculate the
excitation gap $\Delta(N)$, which characterizes the odd-even
oscillations of two-component Fermi systems, as a function of $N$.
For homogeneous two-component Fermi systems with equal masses, the
excitation gap $\Delta$, which equals half the energy it takes to
break a pair, is quite well understood. In the weakly interacting
BCS regime, the excitation gap $\Delta$ becomes exponentially
small~\cite{eagl69,legg80}, indicating vanishingly little pairing.
In the deep BEC regime, on the other hand, the excitation gap
approaches half the binding energy of the free-space dimer,
indicating essentially complete pairing: By adding an extra particle
to the odd $N$ system, the energy of the total system changes by
approximately the binding energy of the free-space dimer. In
addition to these limiting cases, the excitation gap of the
equal-mass two-component Fermi system has been determined throughout
the crossover regime by the FN-DMC
method~\cite{carl03,chan04,carl05}. For unequal-mass systems, in
contrast, the behavior of the gap is much less studied and
understood~\cite{wu06,iski06,linPRA06,iski07,parish2007ftp}.

To define the excitation gap $\Delta(N)$ for trapped unequal-mass
systems, we set $N=2n+1$ and assume $N$ to be odd. The unequal-mass
system is characterized by two chemical potentials, the chemical
potential $\mu_1(N)$ for species one and the chemical potential
$\mu_2(N)$ for species two (see, {\it e.g.}, Ref.~\cite{ring80}),
\begin{eqnarray}
\label{eq_chemup}
E(n+1,n) = E(n,n) + \mu_1(2n+1) + \Delta(2n+1)
\end{eqnarray}
and
\begin{eqnarray}
\label{eq_chemdown}
E(n,n+1) = E(n,n) + \mu_2(2n+1) + \Delta(2n+1).
\end{eqnarray}
Here, $\Delta(2n+1)$ denotes the excitation gap. If $\Delta(2n+1)$
vanishes--- as is the case for the normal system---, then
Eqs.~(\ref{eq_chemup}) and (\ref{eq_chemdown}) reduce to the
``usual'' chemical potentials. Furthermore, $\mu_1(N)$ and
$\mu_2(N)$ coincide for equal-mass systems. To determine $\mu_1(N)$,
$\mu_2(N)$ and $\Delta(N)$, we need an additional relationship. In
condensed matter physics, one typically considers the average of the
two chemical potentials,
\begin{eqnarray}
\label{eq_chemsum}
\frac{1}{2} \left[ \mu_1(2n+1) + \mu_2(2n+1) \right] = \nonumber \\
\frac{1}{2} \left[ E(n+1,n+1)-E(n,n) \right].
\end{eqnarray}
Since the average chemical potential is defined in terms of the
energy of the next smaller and the next larger balanced systems, it
is independent of the odd-even oscillations.
Equations~(\ref{eq_chemup}) through (\ref{eq_chemsum}) can be solved
for $\mu_1(2n+1)$, $\mu_2(2n+1)$ and $\Delta(2n+1)$,
\begin{eqnarray}
\label{eq_chemup2}
\mu_1(2n+1) = \frac{E(n+1,n+1)-E(n,n+1)}{2} \nonumber \\
+ \frac{E(n+1,n)-E(n,n)}{2},
\end{eqnarray}
\begin{eqnarray}
\label{eq_chemdown2}
\mu_2(2n+1) = \frac{E(n+1,n+1)-E(n+1,n)}{2} \nonumber \\
+ \frac{E(n,n+1)-E(n,n)}{2},
\end{eqnarray}
and
\begin{eqnarray}
\label{eq_gap}
\Delta(2n+1) = \frac{E(n+1,n)+E(n,n+1)}{2}\nonumber \\
 - \frac{E(n,n)+E(n+1,n+1)}{2}.
\end{eqnarray}
Note that the energies $E(n+1,n)$ and $E(n,n+1)$ are equal for equal
masses. The excitation gap $\Delta(N)$ and the chemical potentials
$\mu_1(N)$ and $\mu_2(N)$ depend on $N$, $\kappa$, $\omega$ and
$a_s$.

Ultimately, one of the goals is to relate the excitation gaps
calculated for the trapped and the homogeneous systems. For equal
masses and equal frequencies, the densities of the two trapped
species overlap fully. Hence, one might expect that the excitation
gaps of the homogeneous and inhomogeneous systems can be related via
the local density approximation (LDA), which predicts that
$\Delta(N)$ scales with $N$ as $N^{1/3}$. Connecting the excitation
gaps for the homogeneous and trapped systems in this way breaks
down, however, if the extra particle sits near the edge of the gas
cloud; this is the region that is poorly described by the LDA.
Indeed, we present some evidence that the extra particle sits for $N
\gtrsim 11$ near the cloud edge. For unequal masses, the connection
between the two excitation gaps becomes even more challenging,
because one now has to first determine whether the trapped system
exhibits phase separation or not~\cite{linPRA06,paan07}.

For the trapped system,
the density mismatch can be quantified by comparing the density
overlap $O_{N_1,N_2}^{\kappa}$,
\begin{eqnarray}
\label{over} O_{N_1,N_2} ^{\kappa} = a_{ho}^3 \int \rho_1(\vec{r})
\rho_2(\vec{r}) d \vec{r},
\end{eqnarray}
of the unequal-mass system with that of the equal-mass system
for a given scattering length $a_s$.
In Eq.~(\ref{over}),
the one-body densities $\rho_i(\vec{r})$ and the oscillator length
$a_{ho}$ depend on $\kappa$.
In the non-interacting limit, the normalized density mismatch
$O_{N_1,N_2}^{\kappa}/O_{N_1,N_2}^{1}$
reduces to $C_{N_1,N_2}^{\kappa}/C_{N_1,N_2}^{1}$.
In this case,
$O_{N_1,N_2}^{\kappa}/O_{N_1,N_2}^{1}$
equals one for all $\kappa$ if $N_1=N_2=1$
(see Table~II of Ref.~\cite{vonstechtbp}). For
larger $N$, however,
$O_{N_1,N_2}^{\kappa}/O_{N_1,N_2}^1$
decreases from 1 to a finite value that is smaller
than one as $\kappa$ varies from one to infinity.
In particular, the Thomas Fermi approximation predicts
$O_{N_1,N_2}^{\kappa}/O_{N_1,N_2}^1=
315\pi/1024\sqrt{2}\approx0.683$ for large non-interacting systems
($N_1=N_2$) with large $\kappa$.
For the
small unequal-mass systems considered in Sec.~\ref{sec_results}, we
find that the density
mismatch for finite $a_s$ is smaller than that for $a_s=0$.

\subsection{Hyperspherical formulation at unitarity}
\label{sec_hyper}

The two-component Fermi gas at unitarity is characterized by a
diverging scattering length, i.e., $1/a_s=0$. In this regime, the
underlying two-body potential, for sufficiently small $R_0$, has no
characteristic length scale, thus leaving only the size of the
system itself. This elimination  of the two-body length scale is the
key to obtaining a number of analytical results; a particularly
appealing framework for deriving these results employs
hyperspherical coordinates. The hyperspherical formulation has been
primarily developed in the context of few-body
systems~\cite{Delves59,Delves60,mace68,Shitikova77,Chapuisat92,Lin95}.
More recently, some properties of Bose and Fermi gases with
essentially arbitrary number of atoms have been explained
successfully within this
formulation~\cite{bohn98,rittenhouse2005cbc,SorensenPRL02}. The
ability to treat both small and large systems on equal footing makes
the hyperspherical formulation particularly suited for studying the
transition from few- to many-body systems.

We define the hyperspherical coordinates by first
separating off the center-of-mass vector $\vec{R}_{CM}$, and by then
dividing the remaining $3N-3$ coordinates into the
hyperradius $R$ and $3N-4$ hyperangles, collectively
denoted by $\Omega$.
The hyperradius $R$ is defined by
\begin{eqnarray}
\label{eq_hyperradiusrel}
\mu_N R^2=\sum_{i=1}^{N_1} m_1 r_i^2+\sum_{i'=1}^{N_2} m_2
r_{i'}^2 - M
R_{CM}^2,
\end{eqnarray}
and can be viewed as a coordinate
that measures the overall size of the system.
Here, $M$ denotes the total mass of the system,
$M= m_1 N_1 + m_2 N_2$, and $\mu_N$ an arbitrary mass scaling
factor. Usually, the value of $\mu_N$ is chosen so that the
hyperradial potential curves $V_{s_\nu}(R)$, defined below, approach
physically motivated asymptotic values as $R \rightarrow \infty$.

In the adiabatic approximation~\cite{mace68}, the relative wave
function $\Psi^{rel}(R,\Omega)$ reduces to
\begin{eqnarray}
\Psi^{rel}(R,\Omega)=R^{-(3N-4)/2}F_{\nu n}(R)\Phi_\nu(R;\Omega).
\label{HRAdia}
\end{eqnarray}
The antisymmetric Pauli correlations are built into the channel
functions $\Phi_\nu(R;\Omega)$ at the outset. In addition, the
$\Phi_\nu(R;\Omega)$ account for a significant fraction of the
two-body correlations of the system. Within the hyperspherical
approximation, the description of the many-body system reduces to
solving a one-dimensional Schr\"odinger equation in the hyperradial
coordinate $R$,
\begin{eqnarray}
\left(-\frac{\hbar^2}{2\mu_N}\frac{d^2}{dR^2}+V_{s_\nu}(R)+
\frac{1}{2}\mu_N \omega^2 R^2\right)F_{\nu n}(R) \nonumber \\
=E_{\nu n}^{rel}F_{\nu n}(R). \label{HR}
\end{eqnarray}
The effective hyperradial
potential $V_{s_\nu}(R)$ includes part of the kinetic energy and
a contribution due to the short-range two-body interactions.

Assuming zero-range interactions, the adiabatic approximation
becomes exact for a subclass of universal states of the unitary
two-component Fermi gas~\cite{wern06}. For these states, the channel
functions $\Phi_{\nu}$ obey specific boundary conditions imposed by
the zero-range pseudopotential and become independent of $R$.
Furthermore, the functional form of the hyperradial potentials
$V_{s_{\nu}}(R)$ can be derived analytically~\cite{wern06,ritt07},
\begin{eqnarray}
\label{HRpot}
V_{s_\nu}(R)=\frac{\hbar^2
s_\nu(s_\nu+1)}{2\mu_N R^2}.
\end{eqnarray}
 The eigen energies of Eq.~(\ref{HR}) are then given by
\begin{equation}
E_{\nu n}^{rel}=\left(s_\nu+2n+\frac{3}{2} \right)\hbar\omega,
 \label{EHR}
\end{equation}
where $n$ is a non-negative integer, and the hyperradial wave
functions $F_{\nu n}(R)$ (not normalized) by
\begin{eqnarray}
\label{eq_fhyper}
F_{\nu n} (R) = R^{s_{\nu}+1} L_n^{(s_{\nu}+1/2)}(R^2/{\cal{L}}^2)
\exp \left(-\frac{R^2}{2 {\cal{L}}^2} \right),
\end{eqnarray}
where ${\cal{L}}$ denotes the oscillator length associated with
$\mu_N$, ${\cal{L}}=\sqrt{\hbar / (\mu_N \omega)}$, and
$L_n^{(s_{\nu}+1/2)}$ the Laguerre polynomial. The total energy
$E_{\nu n}$ is obtained from $E_{\nu n}^{rel}$ by adding the center
of mass energy. The spacing between states labeled by the same $\nu$
is $2\hbar\omega$ and is thus independent of $s_\nu$. This implies
that knowledge of the lowest eigenenergy $E_{\nu 0}^{rel}$ in each
hyperradial potential curve determines the entire energy spectrum.
This property of the spectrum has also been shown using the scale
invariance properties of unitary systems~\cite{cast04}.
 Transitions
between vibrational levels that lie within a given hyperradial
potential curve $V_{s_{\nu}}(R)$ can be driven by an excitation
operator that depends on $R$ only. Such a driving field results in a
ladder of excitation frequencies of the form $2 k \hbar \omega$,
where $k$ denotes an integer. On the other hand, transitions between
states living in different hyperradial potential curves (labeled by
$\nu$ and $\nu'$) require the driving field to depend on $\Omega$,
or stated more generally, the excitation operator must not commute
with the fixed-hyperradius Hamiltonian. The corresponding excitation
frequencies are, in general, non-integer multiples of $2 \hbar
\omega$ and depend on the difference between $s_{\nu}$ and
$s_{\nu'}$. Thus, knowledge of the entire excitation spectrum
requires determining all $s_{\nu}$. Moreover, the coefficients
$s_\nu$ of the three-body system play a role in determining the
three-body recombination rate for large and negative
$a_s$~\cite{dinc05}, and the lifetime of weakly bound dimers for
large and positive $a_s$~\cite{dinc05}. Similarly, one may expect
that the $s_\nu$ of larger systems play a role in determining the
corresponding quantities for larger systems. Section~\ref{sec_exc}
presents evidence of the $2\hbar\omega$ energy spacing and
determines the $s_\nu$ coefficients for the four-particle system for
various mass ratios.

Equation~(\ref{eq_hyperradiusrel}) defines
the hyperradius $R$ without the CM motion.
Alternatively, we can define a hyperradius $R'$,
\begin{eqnarray}
\label{eq_hyperradiuscm}
M R'^2=\mu_N R^2 + M R_{CM}^2,
\end{eqnarray}
which includes the CM motion
and
represents the rms radius of the system. In the adiabatic
approximation, the total wave function $\Psi(R',\Omega')$ can be
written in terms of the new hyperradius $R'$ as
$\Psi(R',\Omega')=R'^{-(3N-1)/2} \bar{F}_{\nu n}(R')
\bar{\Phi}(R';\Omega')$, where $\Omega'$ collectively denotes the
$3N-1$ hyperangles. Equations~(\ref{HR}) and (\ref{HRpot}) remain
valid if $R$, $\mu_N$ and $F_{\nu n}$ are replaced by $R'$, $M$ and
$\bar{F}_{\nu n}$, respectively. The eigen values of the hyperradial
Schr\"odinger equation equal the eigenenergies $E_{\nu n}$ of the
total system. Defining $x=R'/R'_{NI}$ and $\epsilon_{\nu n}=E_{\nu
n}/E_{NI}$, the hyperradial Schr\"odinger equation can be rewritten
as
\begin{eqnarray}
\left(-\frac{1}{2\mu_{eff}}\frac{d^2}{dx^2}+\frac{s_\nu(s_\nu+1)}{2\mu_{eff}
x^2}+
\frac{1}{2} x^2\right)\bar{F}_{\nu n}(x) \nonumber \\
=\epsilon_{\nu n}\bar{F}_{\nu n}(x), \label{Hx}
\end{eqnarray}
where $\mu_{eff}=E_{NI}^2/(\hbar\omega)^2$. Above, $R'_{NI}$ denotes
the rms radius of the non-interacting system; it can be, using the
virial theorem~\cite{thomasvirialtheorem,wern06}, expressed in terms
of the energy $E_{NI}$ of the non-interacting two-component Fermi
gas,
\begin{eqnarray}
\label{eq_rnihyper}
R'_{NI}=\sqrt{\langle R'^2\rangle_{NI}}=
\sqrt{\frac{\hbar}{M\omega}}\sqrt{\frac{E_{NI}}{\hbar \omega}}.
\end{eqnarray}
The dimensionless coefficients $\bar{C}_N$,
\begin{eqnarray}
\label{Cnbar}
\bar{C}_{N}=\frac{s_0(s_0+1)}{\mu_{eff}}=
\frac{s_0(s_0+1)\hbar^2\omega^2}{E_{NI}^2},
\end{eqnarray}
characterize the ground state of the system at unitarity. The scaled
hyperradius $x$ and the scaled energies $\epsilon_{\nu n}$ remain
finite in the large $N$ limit and are thus particularly well suited
to discuss the large $N$ limit (see Sec.~\ref{sec_unitary}). For
small systems, in contrast, some properties of the system can be
highlighted more naturally using the unscaled hyperradius $R$ or
$R'$.

The coefficients $s_\nu$ describe both the trapped and free systems,
and can be related to the universal parameter $\xi$ of the
homogeneous system~\cite{blumePRL07}. The hyperspherical framework
thus connects few- and many-body quantities and allows one to bridge
the gap between atomic and condensed matter physics.

\section{Numerical techniques}
\label{sec_method}

\subsection{Correlated Gaussian approach}
\label{sec_cg}

The CG method has proven capable of providing an accurate
description of trapped few-body systems with short-range
interactions~\cite{stech07,vonstechtbp, blumePRL07}. The CG method
expands the many-body wave function $\Psi$ in terms of a set of
basis functions $\Phi_{\{d_{ij}\}}$,
\begin{equation}
\Psi
(\vec{r}_1,\cdots,\vec{r}_N) =
\sum_{\{d_{ij}\}} C_{\{d_{ij}\}}\,\Phi _{\{d_{ij}\}}(\vec{r}_1,\cdots,\vec{r}_N),
  \label{TotalWF}
\end{equation}%
where the $C_{\{d_{ij}\}}$ denote expansion coefficients and the
${\{d_{ij}\}}$ a set of widths. Each basis function has the form:
\begin{equation}
\Phi
_{\{d_{ij}\}}
=\mathcal{S}\left\{\psi_0(\vec{R}_{CM})\exp\left(-\sum_{j>i=1}^N
r_{ij}^2/(2d_{ij}^2)\right)\right\}.
  \label{BF}
\end{equation}%
Here, $\psi_0$ is the ground state wavefunction associated with the
center-of-mass vector $\vec{R}_{CM}$, and the operator $\mathcal{S}$
ensures that the basis functions have the proper symmetry under
exchange of two fermions of the same species. Due to the simplicity
of the basis functions, the elements of the Hamiltonian and overlap
matrices can be calculated
analytically~\cite{singer1960uge,suzuki1998sva}. Since the basis
functions depend only on the center of mass vector and the
interparticle distances, i.e., Gaussians centered around $r_{ij}=0$,
the resulting eigenenergies correspond to eigenstates with zero
relative angular momentum $L_{rel}$ and zero total angular momentum
$L$; throughout this work, we do not consider center-of-mass
excitations so that $L_{rel}=L$ for all systems investigated. To
determine the eigenenergies of states  of the $N$-atom system with
non-vanishing $L_{rel}$, we add a spectator atom and solve the
Schr\"odinger equation for the $(N+1)$-atom system. The extra
particle does not interact with the rest of the system but can have
non-vanishing angular momentum. This trick allows us to describe
non-zero angular momentum states of the $N$-atom system. We find,
e.g., that the ground state of the equal-mass three- and five-particle
systems at unitarity has $L_{rel}=1$.

To illustrate how the energies calculated by the CG method converge
with respect to the size of the basis set, we consider the
three-body system with $L=0$ at unitarity.
\begin{figure}[h]
\includegraphics[scale=0.6]{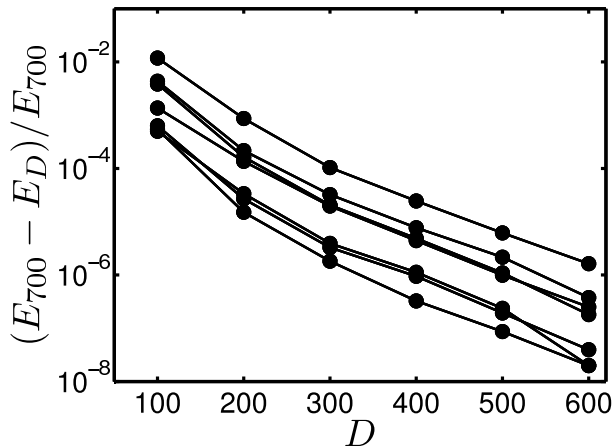}
\caption{Convergence of the energetically lowest-lying
energies as a function of the size $D$ of the basis set
for $N=3$ ($L=0$) at unitarity. The range $R_0$ is fixed at $0.01a_{ho}$.
Solid lines connect the CG energies
(filled circles) of a given state for ease of viewing.}
\label{En3basis}
\end{figure}
We define $E_D$ as the eigenenergies obtained for an optimized basis
set of size $D$. The optimization of the basis functions for a given
size $D$ is performed using the basic ideas of the stochastic
variational approach \cite{suzuki1998sva}. The size of the basis set
is then increased and the new basis functions are optimized.
Figure~\ref{En3basis} shows an example of the convergence of the
lowest few eigenenergies for $R_0=0.01a_{ho}$ as a function of $D$.
The largest $D$ considered in this study is $700$, and the energies
have been tested and are approximately converged for this $D$ value.
Thus, Fig.~\ref{En3basis} shows the normalized difference between
$E_{700}$ and $E_{D}$ for the lowest few eigenenergies.
Figure~\ref{En3basis} shows that the basis set can be improved
systematically.

For larger number of particles, the size of the basis set needs to
be increased. For $N=5$ and $6$, the size of the basis set is
increased up to approximately $D=10^4$. The $N=6$ energies reported
in Ref.~\cite{blumePRL07}, e.g., are calculated for $D= 1.6\times
10^4$. Here, we analyze the convergence of these energies as a
function of $1/D$. Since the energies behave approximately linearly
as a function $1/D$, we can extrapolate straightforwardly to the
limit $D\rightarrow \infty$. The extrapolated energies for $\nu=0$
are $E_{00}=8.48\hbar\omega$, $E_{01}=10.50\hbar\omega$ and
$E_{02}=12.50\hbar\omega$. $E_{00}$ and $E_{01}$ agree with those
reported in Ref.~\cite{blumePRL07} for $D=1.6 \times10^4$ while
$E_{02}$ is only $0.02 \hbar \omega$ lower than the previously
reported value. For $\nu=1$ and $2$, the extrapolated energies are
$E_{10}=10.43\hbar\omega$ and $E_{20}=10.99\hbar\omega$; these
energies are lower by $0.01\hbar \omega$ than those reported in
Ref.~\cite{blumePRL07}. While the extrapolated energies are most
likely closer to the exact eigenenergies than the energies
calculated for $D=1.6\times 10^4$, we note that the extrapolated
energies are no longer variational, i.e., they no longer provide
upper bounds to the exact eigenenergies. Our analysis of the
$\nu\le2$ excited energies shows that the extrapolated energies
follow the expected $2 \hbar \omega$ spacing more closely than those
calculated for the largest $D$ considered, suggesting that the
extrapolation procedure is indeed justified.

In general, the convergence of the energies with respect to the
basis set depends on the scattering length $a_s$ and the number of
states considered. Usually, an accurate determination of the
spectrum at unitarity requires a larger basis than the determination
of the spectrum
on both the weakly-interacting BEC and BCS sides. For equal-mass
systems, a converged basis at unitarity usually describes the
spectrum in the entire crossover region accurately. Of the
equal-mass systems treated, the $N=5$ ($L=1$) calculations have been
the hardest to converge. For $L=1$ states, we can estimate the
uncertainty of the calculations by monitoring the energy of the
spare non-interacting particle, which is known analytically. For
example, for the $N=5$ equal-mass calculations presented in Ref.
\cite{blumePRL07}, the energies of the spare non-interacting
particle deviate from the exact solution by approximately
$0.01\hbar\omega$, which is less than $1\%$. We find that systems
with large $\kappa$ are typically harder to converge than the
corresponding equal-mass systems.

To analyze the effects of finite range interactions we study the
eigenenergies of the three-particle system at unitarity ($L=0$) as a
function of the range $R_0$.
\begin{figure}[h]
\includegraphics[scale=0.6]{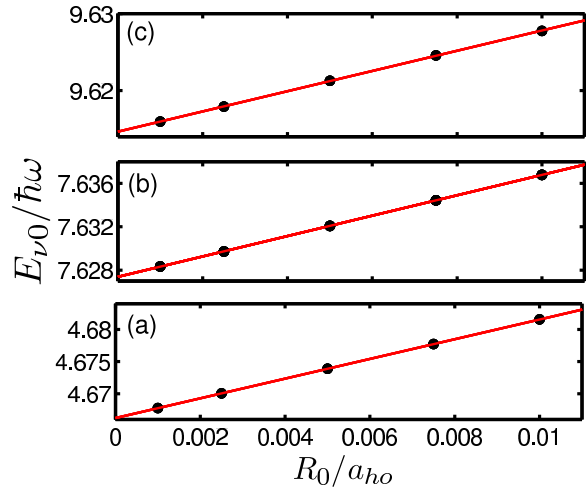}
\caption{(Color Online) Three-body energies $E_{\nu 0}$ at unitarity
for $L=0$ [(a) $\nu=0$, (b) $\nu=1$   and (c) $\nu=2$] as a function
of the range $R_0$. Symbols show the CG energies and solid lines the
linear extrapolation to the $R_0=0$ limit.} \label{En3range}
\end{figure}
Figures~\ref{En3range}(a) through (c) show the energies for the
lowest state in the hyperradial potential curve $V_{s_{\nu}}(R)$
with $\nu=0,1$ and 2. The energies show a linear dependence on
$R_0$, and can thus be extrapolated straightforwardly to the zero
range limit. Neglecting the basis set error, which is estimated to
be smaller than the uncertainty of the extrapolation, we find
$E_{00}=4.66622(1) \hbar \omega$, $E_{10}=7.62738(2) \hbar \omega$,
and $E_{20}=9.61466(4) \hbar \omega$. Our three-body energies
compare favorably with those calculated using the $s_\nu$
coefficients, $\nu=0$ and 1, determined by Ref.~\cite{dinc05} in
Eq.~(\ref{EHR}), $E_{00}=4.6662220 \hbar \omega$ and
$E_{10}=7.6273521 \hbar \omega$. Section~\ref{sec_energyresult}
reports three-particle energies for equal masses for
$R_0=0.01a_{ho}$, which--- according to Fig.~\ref{En3range}--- agree
to better than $0.02 \hbar \omega$ with those calculated in the
zero-range limit. We additionally performed systematic studies of
the dependence of the energies on the range $R_0$ for the three-body
system with equal and unequal masses in the weakly-interacting
molecular BEC regime, where two-body bound states form (see
Secs.~\ref{sec_energy} and \ref{sec_energyresult}), and for the
four-body system. For the five- and six-body calculations, it is
prohibitively expensive to perform calculations for different $R_0$.
For these systems, we estimate the finite range effects based on our
findings for the $N=3$ and 4 systems.

In addition to the energies, Sec.~\ref{sec_struc1} reports
structural properties calculated by the CG approach. The one-body
density and the pair-distribution functions are extracted from the
total wave function $\Psi$ calculated by the CG approach by
integrating $\Psi^2$ over the relevant Jacobi coordinates.

\subsection{Fixed-node diffusion Monte Carlo approach}
\label{sec_fndmc}

For larger systems, the CG approach in our current implementation
becomes prohibitively expensive and we instead determine
first-principles solutions of the time-independent Schr\"odinger
equation using Monte Carlo techniques.

In this study, we use the FN-DMC method~\cite{reyn82,hamm94}, a
variant of the diffusion Monte Carlo (DMC) method, to determine
solutions for up to $N=30$ fermions. The DMC method, which
interprets the system's wave function as a density, allows for the
accurate determination of the energy of nodeless ground states but
is not suited to determine the energy of excited states of bosonic
systems or of fermionic systems. To treat systems whose
eigenfunctions have nodes, the DMC algorithm has to be modified
slightly. Here, we adopt the FN-DMC method, which obtains a solution
of the Schr\"odinger equation that has the same symmetry as a
so-called guiding function $\psi_T$. The FN-DMC method provides, to
within statistical uncertainties, an upper bound to the exact eigen
energy of the many-boson or many-fermion system, i.e., to the
lowest-lying state with the same symmetry as $\psi_T$.

If the nodal surface of $\psi_T$ coincides with that of the
exact eigenfunction, then the FN-DMC method results in the exact
eigen energy of the system. In general, however, the nodal surface
of the exact eigenfunction is not known and the FN-DMC results
depend crucially on the quality of the nodal surface of $\psi_T$. In
this work, we consider three different parametrizations of the nodal
surface of two-component Fermi systems.

The guiding function $\psi_{T1}$ reads
\begin{eqnarray}
\label{eq_t1} \psi_{T1}= \prod_{i=1}^{N_1} \Phi_1(\vec{r}_i) \times
\prod_{i'=1}^{N_2} \Phi_2(\vec{r}_{i'}) \times
F_{node}^{T1}(\vec{r}_1,\cdots,\vec{r}_{N_2})  \times
\nonumber \\
\prod_{i<j}^{N_1} g_{11}(r_{ij}) \times \prod_{i'<j'}^{N_2}
g_{22}(r_{i'j'}) \times \prod_{i,i'}^{N_1,N_2} g_{12}(r_{ii'}).
\end{eqnarray}
The function $F_{node}^{T1}$ determines the nodal structure of
$\psi_{T1}$ and is, for even $N$ and $N_1=N_2$, constructed by
anti-symmetrizing a product of pair functions $f$~\cite{astr04c},
\begin{eqnarray}
\label{eq_t1node1} F_{node}^{T1}= {\cal{A}}
(f(r_{11'}),f(r_{22'}),\cdots, f(r_{N_1 N_2})),
\end{eqnarray}
where ${\cal{A}}$ is the antisymmetrization operator. The pair function $f$ is
given by the free-space two-body solution~\cite{astr04c}: $f$
coincides with the free-space two-body bound state solution for
positive scattering length $a_s$, and with the free-space scattering
solution, calculated at the scattering energy $E_{rel}$, for
negative $a_s$. For $N=6$, we treat $E_{rel}$ as a variational
parameter and find a reduction of the energy of 1 or 2\% for a
finite $E_{rel}$ compared to $E_{rel}=0$. For larger $N$, we simply
use $E_{rel}=0$. For odd $N$, we add a single particle orbital
$\phi_{nl}$ in Eq.~(\ref{eq_t1node1}) so that $F_{node}^{T1}$
becomes, for $N_1=N_2+1$~\cite{bouc88,carl03},
\begin{eqnarray}
\label{eq_t1node2}
F_{node}^{T1}=\nonumber \\
{\cal{A}} (f(r_{11'}),\cdots, f(r_{N_1-1,N_2}),
\phi_{nl}(\vec{r}_{N_1}/a_{ho}^{(1)}))=
\nonumber \\
\mbox{det} \left|
\begin{array}{cccc}
f(r_{11'}) & \cdots & f(r_{1N_2}) & \phi_{nl}(\vec{r}_1/a_{ho}^{(1)}) \\
f(r_{21'}) & \cdots & f(r_{2N_2}) & \phi_{nl}(\vec{r}_2/a_{ho}^{(1)}) \\
\vdots & & \vdots & \vdots \\
f(r_{N_11'}) & \cdots & f(r_{N_1N_2}) &
\phi_{nl}(\vec{r}_{N_1}/a_{ho}^{(1)})
\end{array}
\right|,
\end{eqnarray}
where $a_{ho}^{(i)}= \sqrt{\hbar/(m_i \omega)}$. We consider a
number of different single particle orbitals $\phi_{nl}$, and
determine the optimal $nl$ values by performing a series of FN-DMC
calculations. For the lowest $n$ and $l$, the orbitals read
$\phi_{00}(\vec{r}/a_{ho}^{(1)}) = 1$,
$\phi_{01}(\vec{r}/a_{ho}^{(1)}) = z/a_{ho}^{(1)}$,
$\phi_{20}(\vec{r}/a_{ho}^{(1)}) = 1-2(r/a_{ho}^{(1)})^2/3$ and
$\phi_{02}(\vec{r}/a_{ho}^{(1)}) = 3
(z/a_{ho}^{(1)})^2-(r/a_{ho}^{(1)})^2$.

In Eq.~(\ref{eq_t1}), the $\Phi_i$ ($i=1$ and 2) denote Gaussian
single particle orbitals that depend on a width parameter $b_i$,
$\Phi_i(\vec{r})= \exp(-r^2/(2b_i^2))$. If $b_i= \sqrt{\hbar/(m_i
\omega)}$, $\Phi_i$ coincides with the ground state orbital of the
harmonic oscillator. The parameters $b_1$ and $b_2$ are optimized
variationally. For even $N$ ($N_1=N_2$) and equal masses, we require
$b_1=b_2$. At unitarity, e.g., we find that the $b_i$ are smaller
than the $a_{ho}^{(i)}$, reflecting the attractive nature of the
interspecies interaction potential. If $b_i=a_{ho}^{(i)}$, the
product $\Phi_i(\vec{r}) \phi_{nl}(\vec{r}/a_{ho}^{(i)})$ equals the
harmonic oscillator wave function
$\phi_{nl0}^{(HO)}(\vec{r}/a_{ho}^{(i)})$.

In Eq.~(\ref{eq_t1}), the pair functions $g_{11}$, $g_{22}$ and
$g_{12}$ are introduced to improve the variational energy and to
additionally ensure that the structural properties calculated at the
VMC and FN-DMC levels agree at least qualitatively. The pair
functions $g_{11}$ and $g_{22}$ allow for the effective repulsion
between equal fermions to be accounted for,
\begin{eqnarray}
g_{ii}(r) = \exp(-p_i r^{-q_i})
\end{eqnarray}
for $i=1$ and $2$. The parameters $p_1$, $p_2$, $q_1$ and $q_2$ are
optimized variationally. For even $N$ and equal masses, we require
$p_1=p_2$ and $q_1=q_2$. The pair function $g_{12}$ is parametrized
in terms of the three variational parameters $t$, $p_{12}$ and
$q_{12}$,
\begin{eqnarray}
g_{12}(r) = 1+ t \exp(-p_{12}r^{-q_{12}}).
\end{eqnarray}
The parameters $t$, $p_{12}$ and $q_{12}$ are optimized under the
constrained that $g_{12} \ge 0$.

The guiding function $\psi_{T1}$ is expected to provide a good
description of the system in the weakly-interacting molecular BEC
regime, where we expect bound dimer pairs to form.
Section~\ref{sec_unitary} shows that this wave function also
provides a good description of the unitary gas for sufficiently
large $N$. This is in agreement with FN-DMC studies for the
homogeneous system~\cite{astr04c}. Since each pair function $f$ has
vanishing relative orbital angular momentum, the total angular
momentum $L$ of $\psi_{T1}$ is 0 for even $N$ and $N_1=N_2$. For odd
$N$, $L$ of $\psi_{T1}$ is determined by the angular momentum of
$\phi_{nl}$, i.e., $L=l$.

In addition to $\psi_{T1}$, we consider the guiding function
$\psi_{T2}$,
\begin{eqnarray}
\label{eq_t2} \psi_{T2}= \prod_{i=1}^{N_1} \Phi_1(\vec{r}_i) \times
\prod_{i'=1}^{N_2} \Phi_2(\vec{r}_{i'}) \times
\nonumber \\
\Psi_{node}^{T2}(\vec{r}_1,\cdots,\vec{r}_{N_2}) \times
\prod_{i,i'}^{N_1,N_2} \bar{f}(r_{ii'}).
\end{eqnarray}
The nodal surface of $\psi_{T2}$ is determined by
$\Psi_{node}^{T2}$, which is defined so that the product
$\prod_{i=1}^{N_1} \Phi_1(\vec{r}_i) \times \prod_{i'=1}^{N_2}
\Phi_2(\vec{r}_{i'}) \times \Psi_{node}^{T2}$ coincides for
$b_i=a_{ho}^{(i)}$ with the wave function of $N$ trapped
non-interacting fermions. Thus, the nodal surface of $\psi_{T2}$
coincides with that of the corresponding non-interacting system. The
pair function $\bar{f}$ coincides with the pair function $f$
introduced above for $r \le R_m$, where $R_m$ is a matching point
determined variationally. For $r > R_m$, $\bar{f}$ is given by
$c_1+c_2 \exp(-\alpha r)$. The parameters $c_1$ and $c_2$ are
determined by the condition that $\bar{f}$ and its derivative be
continuous at $r=R_m$ while $\alpha$ is optimized variationally.

The guiding function $\Psi_{T2}$ is expected to provide a good
description of the system in the weakly-interacting BCS regime. In
this regime, we construct the guiding function so that its angular
momentum agrees with that predicted analytically (see
Table~\ref{table1}). Section~\ref{sec_unitary} shows that the
guiding function $\Psi_{T2}$ also provides a good description of
small fermionic systems at unitarity.

Finally, the guiding function $\psi_{T3}$ is constructed following
Eqs.~(3) and (4) of Ref.~\cite{chan07}. We find that $\psi_{T3}$
gives the lowest energy for $N=11$.

Expectation values $\langle A \rangle$ of operators $A$ that do not
commute with the Hamiltonian cannot be calculated as
straightforwardly by the FN-DMC method as the energy. Here, we use
the mixed estimator $\langle A
\rangle_{mixed}$~\cite{whit79,hamm94},
\begin{eqnarray}
\label{eq_mixed}
\langle A \rangle_{mixed} = 2 \langle A \rangle_{DMC} -
\langle A \rangle_{VMC}.
\end{eqnarray}
In Eq.~(\ref{eq_mixed}), $\langle A \rangle_{VMC}$ denotes the
expectation value calculated by the VMC method and $\langle A
\rangle_{DMC}$ that calculated by the FN-DMC method. We note that
some algorithms for the calculation of pure estimators
exist~\cite{barn91,casu95} but we do not use them in this work.

In some cases, we optimize the variational parameters, collectively
denoted by $\vec{p}$, by not only minimizing the energy expectation
value but by additionally ensuring that $\psi_T$ captures selected
structural properties correctly. To this end, we compare the
structural properties calculated by the VMC method for a given
$\vec{p}_0$ with those obtained by the FN-DMC method, which uses
$\psi_T(\vec{p}_0)$ as a guiding function, and then choose a new
parameter set $\vec{p}_1$ so that the VMC structural properties
calculated using $\vec{p}_1$ agree better with the FN-DMC structural
properties calculated using $\vec{p}_0$. This procedure is repeated
till the VMC and FN-DMC structural properties and energy expectation
values agree sufficiently well. For equal-mass systems with $N \le
20$, our VMC energies are at most 15\% higher than the corresponding
FN-DMC energies. The optimization strategy employed here is similar
in spirit to that discussed in Ref.~\cite{chan05} for the
homogeneous system.

\section{Results}
\label{sec_results}

\subsection{Ground state energy in the crossover regime}
\label{sec_energyresult} This section discusses the behavior of the
crossover curve and the excitation gap for $N=3$ for different mass
ratios $\kappa$. This odd $N$ study complements our earlier results
for even $N$~\cite{vonstechtbp}. Our analysis for $N=4$ showed that
the crossover curve $\Lambda_N^{(\kappa)}$ is independent of the
details of the two-body potential and allowed us to extract the
dimer-dimer scattering length $a_{dd}$ and the dimer-dimer effective
range $r_{dd}$ as a function of $\kappa$. The $a_{dd}$ and $r_{dd}$
results from Ref.~\cite{vonstechtbp} are summarized in
Table~\ref{tablenew}. Furthermore, for larger even $N$ systems, we
determined the validity regimes of the analytically calculated
limiting behaviors in the weakly-interacting molecular BEC and
atomic BCS regimes. Our even $N$ study resulted in a deeper
understanding of some of the peculiarities of trapped systems and
emphasized similarities and differences between the trapped and
\begin{table}
\caption{\label{tablenew} Dimer-dimer scattering length $a_{dd}$ and
dimer-dimer effective range $r_{dd}$ obtained using (a) the CG
spectrum and (b) the FN-DMC energies. The reported uncertainties
reflect the uncertainties due to the fitting procedure; the
potential limitations of the FN-DMC method to accurately describe
the energetically lowest-lying gas-like state, e.g., are not
included here (see Sec.~IIIB of Ref.~\cite{vonstechtbp}).}
\begin{ruledtabular}
\begin{tabular}{c|ll|ll}
  $\kappa$ & $a_{dd}/a_s$ (a) & $a_{dd}/a_s$ (b)& $r_{dd}/a_s$ (a)& $r_{dd}/a_s$ (b)\\
\hline
1&   0.608(2) &0.64(1) & 0.13(2)&0.12(4)\\
4& 0.77(1) & 0.79(1) & 0.15(1)& 0.23(1)   \\
8& 0.96(1)  & 0.98(1) & 0.28(1) & 0.38(2)  \\
12& 1.10(1) & 1.08(2) & 0.39(2)& 0.55(2)\\
16& 1.20(1) & 1.21(3) & 0.55(2)& 0.60(5) \\
20& 1.27(2) & 1.26(5)& 0.68(2) & 0.74(5) \\
\end{tabular}
\end{ruledtabular}
\label{EnU4}
\end{table}
homogeneous systems.

The behavior of odd $N$ systems is rich and, in many cases,
qualitatively different from that of even $N$ systems. One
characteristic of odd $N$ systems is the possible change of the
angular momentum of the ground state as the scattering length is
tuned through the BEC-BCS crossover region (see
Sec.~\ref{sec_energy} and Refs.~\cite{kestnerPra07,StetcuCM07}).
\begin{figure}[h]
\includegraphics[scale=0.6]{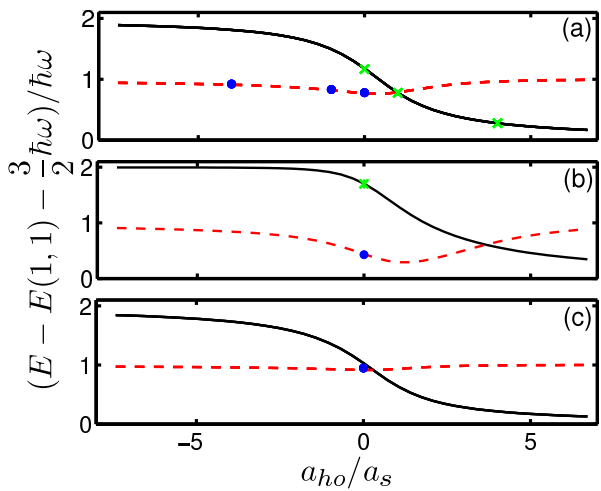}
\caption{(Color Online) Normalized energy $(E-E(1,1)-3\hbar
\omega/2)/\hbar \omega$ for $N=3$ as a function of $a_{ho}/a_s$
calculated by the CG approach (lines). $E$ denotes the three-body
energy for $L=0$ (solid lines) and for $L=1$ (dashed lines): (a)
equal-mass atoms [$\kappa=1$, $E=E(2,1)=E(1,2)$], (b) two heavy
atoms and one light atom [$\kappa=4$, $E=E(2,1)$], and (c) two light
atoms and one heavy atom [$\kappa=4$, $E=E(1,2)$]. The normalized
energy crossover curve $\Lambda_3^{(\kappa)}$,
Eq.~(\protect\ref{eq_cross}), coincides with the dashed and solid
lines, respectively, depending on whether the three-particle ground
state has $L=1$ or $0$. In the CG calculations, the range $R_0$ of
the two-body potential is fixed at $0.01 a_{ho}$. For comparison,
crosses and circles show selected FN-DMC energies for $L=0$ and
$L=1$, respectively. }\label{Ecros3}
\end{figure}
Figure~\ref{Ecros3} shows the three-particle energy $E$, with the
energy $E(1,1)+3\hbar\omega/2$ subtracted, for $L=0$ (solid lines)
and $L=1$ (dashed lines). The upper panel shows results for
$\kappa=1$, and the two lower panels for $\kappa=4$ [panels~(b) and
(c) consider the three-particle system with a spare heavy and a
spare light particle, respectively]. The ground state has $L=1$ for
$a_{ho}/a_s \rightarrow -\infty$ and $L=0$ for $a_{ho}/a_s
\rightarrow \infty$, independent of $\kappa$ and independent of
whether the spare particle is heavy or light. For equal masses, the
change of symmetry occurs at $a_s \approx a_{ho}$. For $\kappa=4$,
in contrast, it occurs at $a_s\approx0.3 a_{ho}$ if the extra
particle is a heavy atom [panel~(b)] and at $a_s\approx3 a_{ho}$ if
the extra particle is a light atom [panel~(c)]. The dashed and solid
lines shown in Fig.~\ref{Ecros3} coincide with the normalized
crossover curve $\Lambda_{N_1,N_2}^{(\kappa)}$,
Eq.~(\ref{eq_cross}), in the region where the ground state of the
three-particle system has $L=1$ and $0$, respectively. The
normalized crossover curve $\Lambda_3^{(\kappa)}$ changes from 1 in
the weakly-interacting molecular BEC regime to 0 in the
weakly-interacting BCS regime.

We find that the normalized $L=1$ energy curve for two heavy atoms
and one light atom [Fig.~\ref{Ecros3}(b)] depends notably on the
range of the underlying two-body potential if the scattering length
$a_s$ is positive. For example, the normalized energy curve changes
by as much as 20\% if the range $R_0$ of the two-body potential
changes from $0.01a_{ho}$ to $0.02a_{ho}$. This comparatively large
dependence on $R_0$ indicates that the properties of the system with
two heavy atoms and one light atom are not fully determined by the
$s$-wave scattering length for the ranges considered. In the $R_0
\rightarrow 0$ limit, the $\kappa=4$ system is expected to behave
universal~\cite{petr05,kartavtsev2007let}. We speculate that the
comparatively strong dependence of the normalized energy curve on
the range for $a_s>0$ is related to the fact that the three-particle
system supports, for sufficiently large $\kappa$, bound states with
negative energy.

For comparison, circles and crosses in Fig.~\ref{Ecros3} show selected
three-particle energies calculated by the FN-DMC method
for $L=0$ and $L=1$, respectively.
The good agreement with the CG results (lines)
indicates that the FN-DMC method can be used to accurately describe
different symmetry states.

Our CG energies for equal-mass systems interacting through
short-range potentials presented in Fig.~\ref{Ecros3}(a) can be
compared with those of Kestner and Duan~\cite{kestnerPra07} obtained
for zero-range interactions. Our $L=1$ energy curve agrees with that
of Kestner and Duan for all scattering lengths $a_s$ considered. The
$L=0$ energy curve, however, only agrees for $a_s<0$. For $a_s>0$,
our results are noticeably lower than those of Kestner and Duan. As
shown below, our $a_s>0$ results for $L=0$ predict the correct
atom-dimer scattering length suggesting that our energies should be
very close to those for $R_0 =0$ and that the disagreement is not
due to finite-range effects. We speculate that the results of
Kestner and Duan might not be fully converged for $a_s>0$ although
other possibilities cannot be excluded.

\begin{figure}[h]
\includegraphics[scale=0.6]{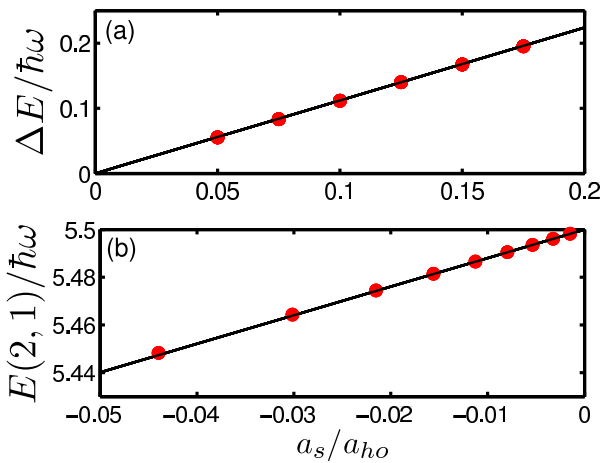}
\caption{(Color Online) Limiting behavior of the ground state energy
for $N=3$ equal mass fermions. (a)  Energy correction $\Delta
E=E(2,1)-E(1,1)-3\hbar \omega/2$ on the BEC side. Circles show the
CG results while the solid line shows the first order correction for
$a_{ad}\approx 1.2 a_s$. (b) Energy $E(2,1)$ on the BCS side.
Circles show the CG results while the solid line shows the first
order correction on the BCS side. }\label{Elim3}
\end{figure}

Figures \ref{Elim3}(a) and (b) present the BCS and BEC limiting
behaviors for an equal mass system with $N=3$. The perturbative
expression, Eq.~(\ref{eq_expbec}), on the BEC side is expected to be
applicable if $R_0 \ll a_s \ll a_{ho}$; thus, we choose a small
$R_0$, i.e., $R_0=0.005a_{ho}$, in the CG calculations. The energy
is in this region determined by the atom-dimer scattering length
$a_{ad}$ [see Eq.~(\ref{eq_expbec})]. The CG energies change
linearly with $a_s$, showing that $a_{ad}$ is proportional to $a_s$,
i.e., $a_{ad}=c_{ad} a_s$. A simple linear fit to the CG results
predicts $c_{ad}\approx 1.21$, in good agreement with previous
studies~\cite{skor56,petr03}, which found $a_{ad}\approx 1.2 a_s$. A
solid line in Fig.~\ref{Elim3}(a) shows the resulting linear
expression. A more sophisticated analysis accounts for the
energy-dependence of $a_{ad}$~\cite{blum02,bold02}, which  results
in a more reliable determination of $c_{ad}$ and also a
determination of the effective range $r_{ad}$~\cite{vonstechtbp}.
Considering the three lowest energy levels on the BEC
side~\cite{vonstechtbp}, we obtain $c_{ad} \approx 1.18(1)$ and
$r_{ad}\approx 0.08(1) a_s$. It was suggested earlier~\cite{petr05a}
that the atom-dimer system is characterized by a soft-core repulsion
with range of the order of $a_s$; our calculations support this
general picture but predict a range about ten times smaller than
$a_s$. On the BCS side, the first order correction varies also
linearly with $a_s$. Circles in Fig. \ref{Elim3}(b) show the CG
results while the solid line shows the prediction from
Eq.~(\ref{eq_expbcs}). Good agreement is observed in both limiting
behaviors.

Our CG energies for $N=2$, $3$ and $4$ can be readily combined to
determine the excitation
gap $\Delta(3)$, Eq.~(\ref{eq_gap}).
\begin{figure}[h]
\includegraphics[scale=0.6]{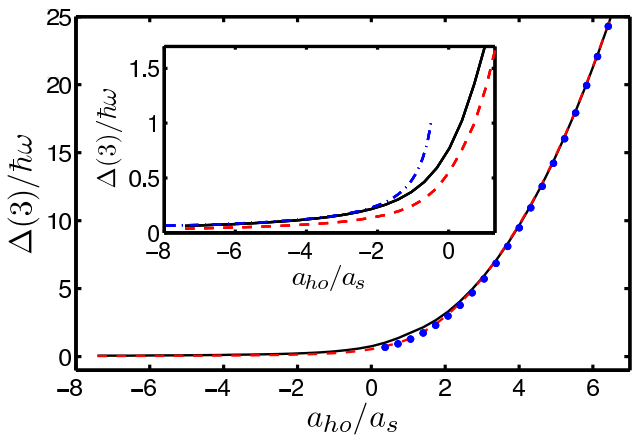}
\caption{(Color Online) Excitation gap $\Delta(N)$ for $N=3$ as a
function of $a_{ho}/a_s$ calculated by the CG approach for
$\kappa=1$ (solid line) and $\kappa=4$ (dashed line). Circles
present the BEC limiting behavior $3\hbar\omega/2-E(1,1)/2$ which is
independent of $\kappa$. The inset shows a blow-up of the region
where $\Delta(3)$ is smallest; in this region, the dependence of
$\Delta(3)$ on $\kappa$ is most pronounced. The dash-dotted line
shows the limiting behavior for $\kappa=1$ obtained by approximating
the $E(N)$ in Eq.~(\ref{eq_gap}) by their perturbative values,
Eq.~(\ref{eq_expbcs}). } \label{gap3cros}
\end{figure}
Figure~\ref{gap3cros} shows the excitation gap $\Delta(3)$ as a
function of $a_{ho}/a_s$ for two different mass ratios, i.e.,
$\kappa=1$ and $4$. In the weakly-interacting molecular BEC regime,
the excitation gap approaches $3\hbar\omega/2-E(1,1)/2$ (circles),
independent of the mass ratio. In the weakly-interacting BCS regime,
however, the excitation gap depends on the mass ratio (see inset of
Fig.~\ref{gap3cros}). For equal masses, $\Delta(3)$ is very well
described by the perturbative expression for $a_s \lesssim -0.5
a_{ho}$ (dash-dotted line in the inset). Figure~\ref{gap3cros} shows
that $\Delta(3)$ is smaller for $\kappa=4$ than for $\kappa=1$.
Intuitively, this might be expected since the radial densities of
the two species do not fully overlap for unequal masses (recall, we
consider the case where species one and two experience the same
trapping frequency). Thus, the pairing mechanism is expected to be
less efficient in the unequal-mass system, especially on the BCS
side, than in the equal-mass system. The next section discusses the
behavior of the excitation gap at unitarity in more detail.

\subsection{Ground state energy at unitarity}
\label{sec_unitary}
This section explores the odd-even behavior of two-component Fermi
gases at unitarity. In particular, we present the excitation gap for
equal-mass systems with up to $N=30$ fermions and interpret the
behaviors of these systems within the hyperspherical framework. We
also discuss the excitation gap for small unequal-mass systems.

Table~\ref{tab1} summarizes selected CG and FN-DMC energies for
small equal-mass systems at unitarity. Some of these energies were
already reported in Refs.~\cite{vonstechtbp,blumePRL07}, and we
include them in Table~\ref{tab1} for comparative purposes.
\begin{table}
\caption{\label{tab1} CG and FN-DMC energies $E$ at unitarity for
small equal-mass systems with angular momentum $L=0$ and 1. The
CG energies are calculated for the Gaussian interaction potential
with $R_0=0.01a_{ho}$ for $N=3$ and $4$, and with $R_0=0.05a_{ho}$
for $N=5$ and 6. The FN-DMC energies are calculated for the square
well interaction potential with $R_0=0.01a_{ho}$. The guiding
functions $\psi_{T1}$ and $\psi_{T2}$, Eqs.~(\ref{eq_t1}) and
(\ref{eq_t2}), are used to obtain the energies of states
with $L=0$ and $1$,
respectively. }
\begin{ruledtabular}
\begin{tabular}{l|c|lll|ll}
 $N$ & $L$ & $E/(\hbar \omega)$ (CG) & $E/(\hbar \omega)$
(FN-DMC) \\ \hline
3 &0& 4.682 & 4.67(3) \\
3 &1& 4.275 & 4.281(4) \\
4 &0& 5.028 & 5.051(9)\\
5 &0& 8.03  & 8.10(3) \\
5 &1& 7.53  & 7.61(1) \\
6 &0& 8.48  & 8.64(3) \\
7 &0&       & 11.85(5)\\
7 &1&       & 11.36(2)\\
8 &0&       & 12.58(3)\\
9 &0&       & 15.84(6)\\
9 &1&       & 15.69(1)\\
\end{tabular}
\end{ruledtabular}
\label{EnU}
\end{table}
A comparison of the CG and FN-DMC energies for $N\le 6$ shows that
the FN-DMC energies agree to within 2\% with the CG energies for
both $L=0$ and $1$ states. This agreement suggests that the nodal
surface used in the FN-DMC calculations is quite accurate. Thus,
Table~\ref{tab1} shows that the FN-DMC method allows not only for an
accurate description of the ground state but also of excited states.
For $N=9$,  the energy of the $L=1$ state is by only about
$0.15\hbar \omega$ smaller than that of the $L=0$ state. The ground
state energies for larger $N$ are reported in Table~II of
Ref.~\cite{blumePRL07}. For both even and odd $N$ ($N>9$), we find
that the angular momentum of the lowest energy state at unitarity is
zero. Our FN-DMC energies thus suggest that the total angular
momentum of the lowest energy states at unitarity has $L=1$ for
small odd $N$ systems and $L=0$ for larger odd $N$ systems. We note
that this conclusion depends crucially on the construction of the
nodal surface entering the FN-DMC calculations. For $N=19$, e.g.,
the energies at unitarity for $L=2$ and 1 are less than $0.8\hbar
\omega$ higher than the $L=0$ energy; thus, the definite
determination of the ordering of the states at unitarity with
different angular momenta remains a challenge for odd-$N$ systems
with $N>9$.

\begin{figure}[h]
\includegraphics[scale=0.6]{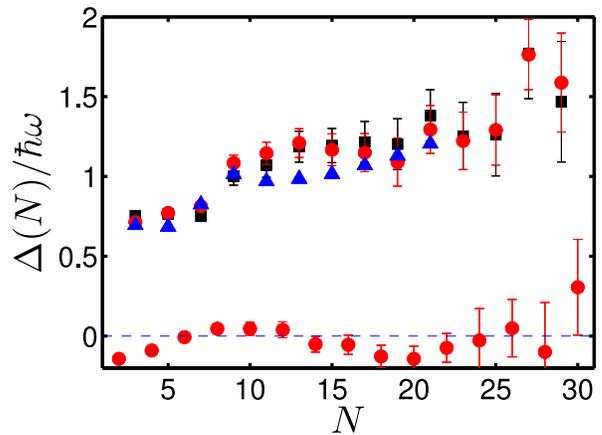}
\caption{(Color Online) Excitation gap $\Delta(N)$ (squares) and
residual energy $E_{00}(N)-E_{fit}$ (circles) for equal-mass Fermi
systems at unitarity as a function of $N$ calculated from the FN-DMC
energies. Triangles show $\Delta(N)$ calculated using density
functional theory \cite{bulgac07}.} \label{gapN}
\end{figure}

For homogeneous systems, the ground state energy per particle at
unitarity $E_u$ is related to the energy per particle $E_{FG}$ of
the non-interacting system by a universal proportionality
constant~$\xi$, $E_u=\xi E_{FG}$~\cite{carl03,astr04c,carl05}.
Applying this result to the trapped unitary system with even $N$
through the LDA, the ground state energy $E_{00}(N)$ of the trapped
system becomes directly proportional to the energy $E_{NI}$ of the
non-interacting trapped system~\cite{vonstechtbp},
\begin{eqnarray}
E_{00}(N)=\sqrt{\xi} E_{NI}.
\end{eqnarray}
An analysis of our FN-DMC energies for $N=2-30$ suggests that the
trapped unitary system shows little shell structure. This motivates
us to ``smooth'' the non-interacting energies, i.e., we approximate
$E_{NI}$ by the extended Thomas-Fermi expression~\cite{brack},
\begin{equation}
E_{NI,ETF} = \hbar \omega \frac{1}{4} (3N)^{4/3}
\left( 1+ \frac{1}{2} (3N)^{-2/3} \right).
  \label{ETF}
\end{equation}%
To determine the proportionality constant $\xi$, we fit our even $N$
energies for $N=2-30$ to the expression $\sqrt{\xi_{tr}}
E_{NI,ETF}$. We find $\xi_{tr}=0.467$, and denote the resulting
energies by $E_{fit}$. This value is in very good agreement with our
previous result, $\xi_{tr}=0.465$, obtained by including the
energies for $N=2-20$ only~\cite{vonstechtbp}. Circles in
Fig.~\ref{gapN} show the residual energy $E_{00}(N)-E_{fit}$ for
both even and odd $N$. For even $N$, the energy difference
$E_{00}(N)-E_{fit}$ is at most $0.15 \hbar \omega$ (except for
$N=30$, for which the error bar is large). This suggests that the
energies of the trapped unitary system are indeed quite well
described by $\sqrt{\xi_{tr}} E_{NI,ETF}$; in other words, our
energies show little residual shell structure.
As expected, the odd $N$ energies are not even quantitatively
described correctly by $E_{00}(N)-E_{fit}$. Instead, Fig.~\ref{gapN}
shows that the residual
energy $E_{00}(N)-E_{fit}$ for odd $N$
(circles) agrees quite well with the excitation gap $\Delta(N)$
(squares). For comparison, triangles in Fig.~\ref{gapN} show the
excitation gap calculated using DFT~\cite{bulgac07}. The good
agreement between the DFT and FN-DMC results is encouraging.

The ground-state energies $E_{00}(N)$ determine
the coefficients $s_0$ [see Eq.~(\ref{EHR})] of the hyperradial potential
$V_{s_0}(R)$ [see Eq.~(\ref{HRpot})].
\begin{figure}[h]
\includegraphics[scale=0.6]{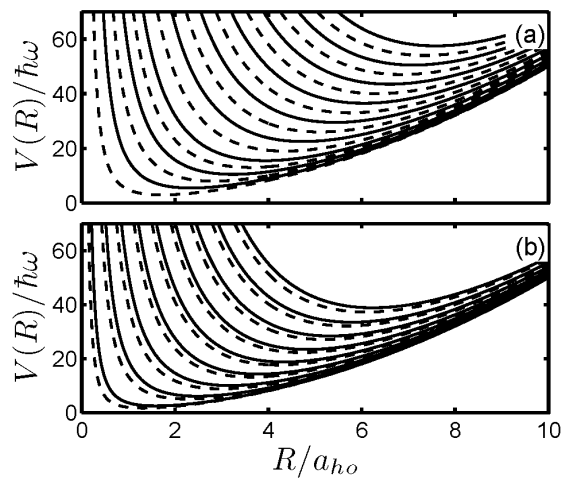}
\caption{ Hyperradial potential curves $V(R)$ for equal-mass
two-component Fermi systems with (a) vanishing interactions and (b)
infinitely strong interactions as a function of $R$. The hyperradial
potential curves naturally appear ordered as $N$ increases: Solid
lines correspond, from bottom to top, to $N=4-20$ ($N$ even), while
dashed lines correspond, from bottom to top, to $N=3-19$ ($N$ odd).
} \label{pc}
\end{figure}
Figures~\ref{pc}(a) and (b) show the lowest hyperradial potential
curves $V(R)$ [$V(R)=V_{s_0}(R)+V_{trap}(R)$, where
$V_{trap}(R)=\frac{1}{2} \mu_N \omega^2 R^2$ and $\mu_N=m$] for
$N=3-20$ in the non-interacting limit and at unitarity,
respectively. The small $R$ behavior of $V(R)$ is dominated by
$V_{s_0}(R)$ while the large $R$ behavior of $V(R)$ is dominated by
$V_{trap}(R)$. Comparison of Figs.~\ref{pc}(a) and (b) shows that
the attractive interactions lead to a lowering of the potential
curves at unitarity compared to those of the non-interacting system.
Furthermore, the $V(R)$ at unitarity appear ``staggered'', i.e.,
odd-even oscillations are visible, reflecting the finite excitation
gap at unitarity. In the non-interacting limit, in contrast, the
excitation gap is zero and no odd-even staggering of the hyperradial
potential curves is visible.

\begin{figure}[h]\includegraphics[scale=0.6]{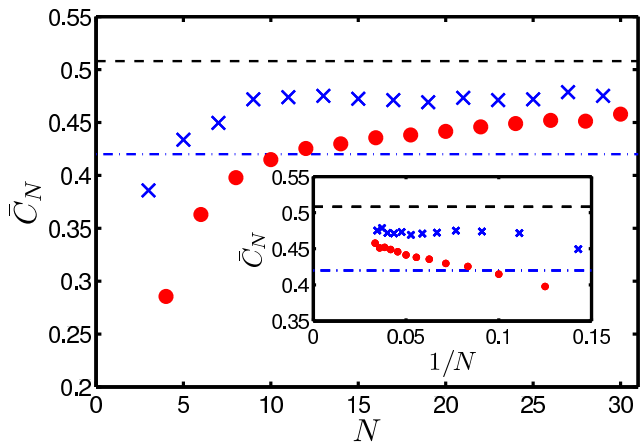}
\caption{(Color Online)  Normalized coefficients $\bar{C}_N$,
Eq.~(\protect\ref{Cnbar}) with $E_{NI}$ replaced by $E_{NI,ETF}$, as
a function of $N$; values for even $N$ are shown by circles and
values for odd $N$ by crosses. The dash-dotted line shows the value
$\xi=0.42$ obtained by FN-DMC calculations for the homogeneous
system~\protect\cite{astr04c,carl05}, while a dashed curve shows the
value $\xi=0.508$ obtained with a renormalization
procedure~\protect\cite{vonstech07}. The inset shows the same
quantities as a function of $1/N$ instead of $N$. } \label{CN}
\end{figure}

To extrapolate to the large $N$ limit, Fig.~\ref{CN} shows the
normalized coefficients $\bar{C}_N$, Eq.~(\ref{Cnbar}) with $E_{NI}$
replaced by $E_{NI,ETF}$, as a function of $N$ (just as in our
analysis of the energies $E_{00}$ we find it useful to smooth the
energies $E_{NI}$).
The coefficients $\bar{C}_N$ oscillate between two smooth curves, a
curve for even $N$ (circles) and a curve for odd $N$ (crosses). As
$N$ increases, the difference between the two curves decreases. In
the large $N$-limit, the value of $\bar{C}_N$ for two-component
Fermi gases at unitarity should approach the universal parameter
$\xi$~\cite{blumePRL07}. This can be shown by relating the ground
state energy obtained within the hyperspherical framework,
Eq.~(\ref{EHR}), to the LDA prediction (see above), or by applying
renormalized zero-range interactions within the hyperspherical
framework~\cite{ritt07}. The dash-dotted and dashed lines in
Fig.~\ref{CN} show the $\xi$ value obtained by FN-DMC calculations
for the homogeneous system ($\xi=0.42$)~\cite{astr04c,carl05} and
the $\xi$ value obtained with a renormalization procedure
($\xi=0.508$)~\cite{vonstech07}, respectively. It is generally
believed that the FN-DMC calculations provide the most reliable
estimate for $\xi$ to date. For comparison, our energies for the
trapped system predict $\xi_{tr}=0.467$ (see above). The circles in
Fig.~\ref{CN} approach this value. We attribute the fact that
$\xi_{tr}$ is larger than the corresponding value of the bulk
system, i.e., $\xi =0.42$, to the comparatively small system sizes
($N\le30$) included in our analysis. If this was true, we would
expect the circles in the main part of Fig.~\ref{CN} to turn around
at larger $N$ values. We note that we cannot rule out that the nodal
surface entering our FN-DMC calculations might not be optimal.

\begin{figure}[h]
\includegraphics[scale=0.6]{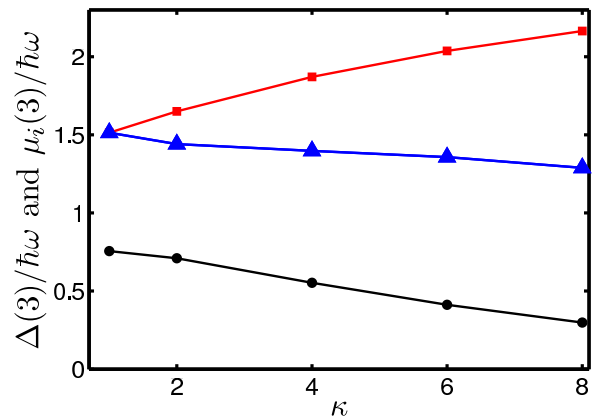}
\caption{(Color Online) Circles show the excitation gap $\Delta(N)$
for $N=3$ as a function of the mass ratio $\kappa$ at unitarity.
Triangles and squares show the chemical potentials $\mu_1(3)$ and
$\mu_2(3)$, respectively.} \label{gap3}
\end{figure}

In addition to equal-mass unitary systems, we study small systems
with unequal masses at unitarity. Figure~\ref{gap3} shows the
excitation gap $\Delta(N)$
for $N=3$ at unitarity as a function of the mass ratio $\kappa$.
$\Delta(3)$ decreases from about $0.8 \hbar \omega$ for $\kappa=1$
to about $0.3 \hbar \omega$  for $\kappa=8$. A decrease of the
excitation gap as a function of $\kappa$ has recently also been
reported for the homogeneous unequal-mass system at
unitary~\cite{astr07}. To better understand the decrease of
$\Delta(N)$ with increasing $\kappa$, triangles and squares in
Fig.~\ref{gap3} show the chemical potentials $\mu_1(3)$ and
$\mu_2(3)$ for the two species. The decrease of $\mu_1$ is related
to the fact that trimers with negative energy form for sufficiently
large $\kappa$. We additionally note that the densities of the light
and heavy particles do not fully overlap. This effect is unique to
the trapped system (the study of the homogeneous system with unequal
masses~\cite{astr07} assumes equal densities of the two components
and full pairing). Simple arguments lead one to conclude that a
partial density overlap as opposed to a full density overlap leads
to a decrease of the excitation gap. Thus, it is not clear if the
decrease of $\Delta(3)$ visible in Fig.~\ref{gap3} with $\kappa$ is
due to the same mechanisms that lead to a decrease of $\Delta$ in
the homogeneous system or due to the specifics of the trapping
potentials, or possibly both.

\subsection{Excitation spectrum at unitarity}
\label{sec_exc} Excitation spectra of two-component Fermi gases are
rich. For four equal-mass fermions, e.g., Ref.~\cite{stech07} shows
how the spectrum evolves from the non-interacting limit for small
$|a_s|$, $a_s<0$, to different families in the small $a_s$ region,
$a_s>0$: One family consists of states that describe two bound
dimers, another consists of states that describe a bound dimer plus
two atoms, and yet another consists of states that describe a gas.
Between these two limiting cases is the unitary region where the
eigenspectrum is expected to be characterized by unique properties,
similar to those of the non-interacting system (see
Sec.~\ref{sec_hyper}). In particular, in the unitary regime families
of eigenenergies separated by $2 \hbar \omega$ are expected to
exist~\cite{wern06}. This prediction has recently been verified for
up to six particles with equal masses to within numerical accuracy,
i.e., to within 2\%~\cite{blumePRL07}. Here we extend our analysis
to unequal-mass systems with $N=4$ and $L=0$.

\begin{figure}[h]
\includegraphics[scale=0.5]{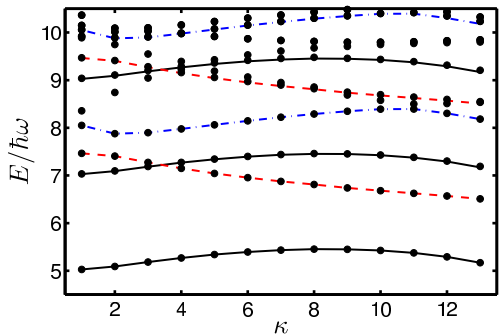}
\caption{(Color Online) Four-body energy spectrum for $L=0$ at
unitarity as a function of $\kappa$. Circles correspond to the
numerical results obtained by the CG approach. Solid, dashed and
dash-dotted lines show the energies $E_{\nu0}+2n\hbar \omega$ for
$\nu=0$, 1 and 2, respectively ($n=0,1,\cdots$). }\label{Spec4Uni}
\end{figure}

Circles in Fig.~\ref{Spec4Uni} show the zero angular-momentum energy
spectrum calculated by the CG approach for four particles at
unitarity as a function of $\kappa$. The range of the Gaussian
potential is $R_0=0.01a_{ho}$.
To analyze the eigenenergies, we employ the hyperspherical
framework. Assuming that the separation of the wave function (see
Sec.~\ref{sec_hyper}) holds for the short-range interactions
considered here, we expect that the energy spectrum consists of
families of energy levels separated by $2 \hbar \omega$. Solid lines
show the energies $E_{0 0}+2n \hbar \omega$ ($n$ non-negative
integer), where $E_{00}$ denotes the lowest positive energy of the
spectrum (for sufficiently large $\kappa$, negative energy states
form; these are not shown in Fig.~\ref{Spec4Uni}). The agreement
between the solid lines and the CG energies indicates that the
$2\hbar\omega$ spacing, predicted for zero-range interactions, is
fulfilled within our numerical accuracy. We repeat this exercise for
the next family of energy levels: Dashed lines show the energy $E_{1
0}+2n \hbar \omega$, where $E_{10}$ corresponds to the lowest
positive energy not yet assigned to a family. Similarly, dash-dotted
lines connect states belonging to the third family. In addition to
the just outlined assignment of quantum numbers, we checked in a few
cases that the hyperradial wave functions $F_{\nu n}(R)$
corresponding to the energies $E_{\nu n}$ possess $n$ hyperradial
nodes (see also Sec.~\ref{sec_struc2}). The lines in
Fig.~\ref{Spec4Uni} show a crossing of energy levels belonging to
different families at $\kappa \approx 4$. In close vicinity to this
crossing, the spacing may not be exactly $2 \hbar \omega$.

As already pointed out in the previous section, the energies
$E_{\nu0}$ determine the coefficients $s_{\nu}$ of the hyperradial
potential curves $V_{s_{\nu}}(R)$. Table~\ref{EnU4} summarizes the
three smallest coefficients for various $\kappa$.
\begin{table}
\caption{\label{tab2} Coefficients
$s_{\nu}$ of the hyperradial potential curves $V_{s_{\nu}}(R)$,
Eq.~(\ref{HRpot}), for the $N=4$ system with $L=0$
for various mass ratios $\kappa$.
}
\begin{ruledtabular}
\begin{tabular}{c|lll||c|lll}
  $\kappa$ & $s_0$ & $s_1$  &$s_2$ & $\kappa$ & $s_0$ & $s_1$  &$s_2$\\
\hline
1& 2.03 &4.46 & 5.05  & 8 & 2.45 &3.81 & 5.29 \\
2& 2.09 &4.41 & 4.88  & 9 & 2.45 &3.74 & 5.35  \\
3& 2.18& 4.27 & 4.90  & 10 & 2.42 &3.68 & 5.39 \\
4& 2.27  &4.15 &4.98  & 11 & 2.37 & 3.62
& 5.39  \\
5& 2.34  &4.04&  5.06 & 12 & 2.29 &3.57 & 5.30 \\
6& 2.40& 3.95 & 5.15 & 13 & 2.17 &3.51 & 5.18\\
7& 2.43 & 3.88 & 5.22  & & & &\\
\end{tabular}
\end{ruledtabular}
\label{EnU4}
\end{table}
To the best of our knowledge, these are the first calculations of
the $s_{\nu}$ for four-particle systems with unequal masses.

\subsection{Structural properties along the BEC-BCS crossover}
\label{sec_struc1}

In addition to the energetics, we analyze the one-body densities and
pair distribution functions of two-component Fermi systems in the
crossover regime for different $\kappa$. While the densities
$\rho_i(\vec{r})$ of $L=0$ states are spherically symmetric, those
of states with $L>0$ are not spherically symmetric. In the
following, we determine the averaged radial densities $\rho_i(r)$,
normalized so that $4\pi \int \rho_i(r) r^2 dr=N_i$; $4\pi r^2
\rho_i(r)/N_i$ tells one the probability of finding a particle with
mass $m_i$ at a distance $r$ from the center of the trap. If
$N_1=N_2$ and $m_1=m_2$, the radial one-body densities $\rho_1(r)$
and $\rho_2(r)$ coincide. If $m_1$ and $m_2$ or $N_1$ and $N_2$
differ, however, the radial one-body densities $\rho_1(r)$ and
$\rho_2(r)$ are, in general, different. We also determine the
averaged radial pair distribution functions $P_{ij}(r)$, normalized
so that $4\pi \int P_{ij}(r) r^2 dr=1$; $4\pi r^2 P_{ij}(r)$ tells
one the probability to find a particle of mass $m_i$ and a particle
of mass $m_j$ at a distance $r$ from each other. For notational
simplicity, we refer to the radial one-body densities as one-body
densities and to the radial pair distribution functions as pair
distribution functions in the following.

\begin{figure}[h]
\includegraphics[scale=0.6]{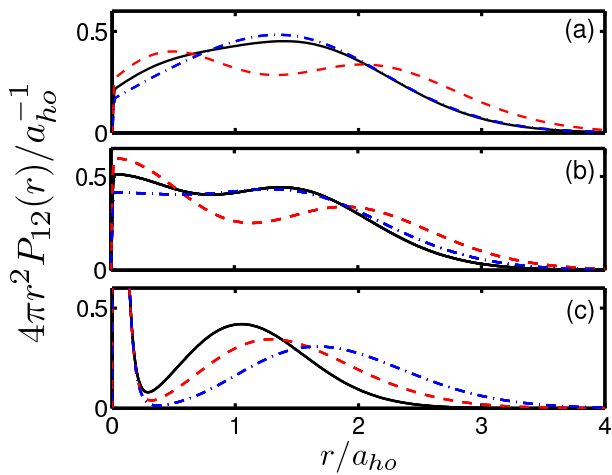}
\caption{(Color Online) Pair distribution functions $P_{12}(r)$,
multiplied by $r^2$, for equal mass two-component Fermi systems with
$N=3$ and $L=0$ (dashed lines), $N=3$ and $L=1$ (dash-dotted lines),
and $N=4$ and $L=0$ (solid lines) obtained by the CG approach for
three different scattering lengths $a_s$: (a) $a_s=-a_{ho}$ (BCS
regime), (b) $1/a_s=0$ (unitarity), and (c) $a_s=0.1a_{ho}$ (BEC
regime). The pair distribution function for $N=4$ and
$a_s=0.1a_{ho}$ [solid line in panel (c)] is shown in more detail in
Fig.~\protect\ref{Corr4BEC}. }\label{Corr3and4}
\end{figure}

Figure~\ref{Corr3and4} shows the pair distribution
function $P_{12}(r)$ for $N=3$ (dash and dash-dotted lines
correspond to $L=0$ and $1$, respectively) and $N=4$ (solid lines)
along the crossover for $\kappa=1$. Panel~(a) shows results for
$a_s=-a_{ho}$, panel~(b) for $1/a_s=0$ and panel~(c) for
$a_s=0.1a_{ho}$. Interestingly, the pair distribution functions for
$N=3$ and 4 show a similar overall behavior. In the BCS regime
[Fig.~\ref{Corr3and4}(a)], the quantity $P_{12}(r)r^2$ shows a
minimum at small $r$ (for very small $r$, $P_{12}(r)r^2$ goes
smoothly but steeply to zero; this rapid change of $P_{12}(r)r^2$ is
hardly visible on the scale shown in Fig.~\ref{Corr3and4}). At
unitarity [Fig.~\ref{Corr3and4}(b)], $P_{12}(r)r^2$ shows a maximum
at small $r$ and a second peak at larger $r$. In the BEC regime
[Fig.~\ref{Corr3and4}(c)], the two-peak structure is notably more
pronounced. The peak at small $r$ indicates the formation of
tightly-bound dimers (one dimer for $N=3$ and two dimers for $N=4$),
while the peak between $1a_{ho}$ and $2a_{ho}$ is related to the
presence of larger atom-atom distances set approximately by the
atom-dimer distance for the three-body system and the dimer-dimer
distance for the four-body system. This interpretation suggests that
the three-particle system has one small and one large interspecies
distance, and the four-particle system has two small and two large
interspecies distances. Indeed, integrating $P_{12}(r)$ for $N=3$
and 4 from $0$ to the $r$ value at which $P_{12}(r)r^2$ exhibits the
minimum, we find that the likelihood of being at small distances
(forming a molecule) and being at large distances is the same.

We now analyze the pair distribution function $P_{12}(r)$ for $N=4$
\begin{figure}[h]
\includegraphics[scale=0.6]{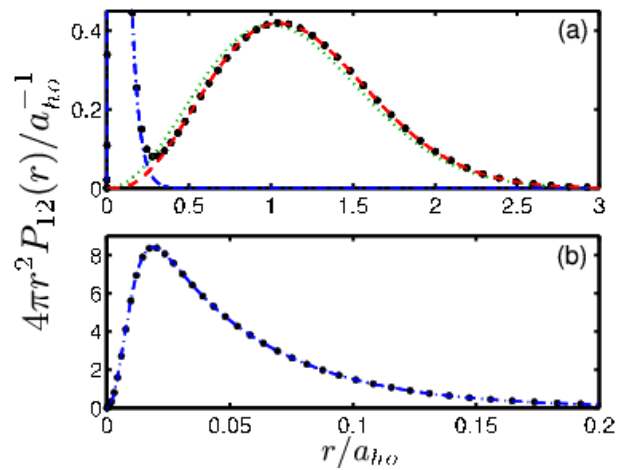}
\caption{(Color Online) (a) Circles show the pair distribution
function $P_{12}(r)$, multiplied by $r^2$, for $a_s=0.1a_{ho}$ (BEC
regime) calculated by the CG approach for $N=4$ and $\kappa=1$
[note, this quantity is also shown by a solid line in
Fig.~\protect\ref{Corr3and4}(c)]. For comparison, the dash-dotted
line (blue online) shows $P_{12}(r)r^2$ for two atoms of mass $m$
with the same scattering length but normalized to $1/2$, the dashed
line (red online) shows $P_{12}(r)r^2$ for two trapped bosonic
molecules of mass $2m$ interacting through a repulsive effective
potential with $a_{dd}=0.6a_s$, and the dotted line (green online)
shows $P_{12}(r)r^2$ for two trapped non-interacting bosonic
molecules of mass $2m$. Panel (b) shows a blow-up of the small $r$
region.\label{Corr4BEC}}
\end{figure}
more quantitatively. Dash-dotted lines in Figs.~\ref{Corr4BEC}(a)
and (b) show the pair distribution function $P_{12}(r)$, multiplied
by $r^2$, for two trapped atoms with $a_s=0.1a_{ho}$ (normalized to
$1/2$). This dimer curve is essentially indistinguishable from the
small $r$ part of the four-particle pair distribution function
(circles). To describe the large $r$ part of the four-particle pair
distribution function, we consider two bosonic molecules of mass
$2m$, which interact through an effective repulsive potential with
dimer-dimer scattering length $a_{dd} \approx
0.6a_s$~\cite{petr05,vonstechtbp}. The dashed line in
Fig.~\ref{Corr4BEC}(a) shows the pair distribution function for this
system under external confinement. This dashed curve is essentially
indistinguishable from the large $r$ part of the pair distribution
function for the four-particle system. For comparison, a dotted line
shows the pair distribution function for two non-interacting trapped
bosons of mass $2m$. Figure~\ref{Corr4BEC} indicates that the
effective repulsive interaction between the two dimers is crucial
for reproducing the structural properties of the four-body system
accurately. Our analysis shows that the entire pair distribution
function $P_{12}(r)$ of the four-body system in the
weakly-interacting molecular BEC regime can be described
quantitatively in terms of a ``dimer picture''.

We now return to Fig.~\ref{Corr3and4} and discuss how the
symmetry-inversion of the $N=3$ system along the crossover (see
Sec.~\ref{sec_energyresult}) is reflected in $P_{12}(r)$. In the BCS
regime and at unitarity [Figs.~\ref{Corr3and4}(a) and (b)],
$P_{12}(r)$ shows less structure for $L=1$ than for $L=0$. In the
weakly-interacting molecular BEC regime [Fig.~\ref{Corr3and4}(c)],
the pair distribution function for $L=0$ nearly coincides with that
for $L=1$ at small $r$ but is more compact than that for $L=1$ at
large $r$.

\begin{figure}[h]
\includegraphics[scale=0.6]{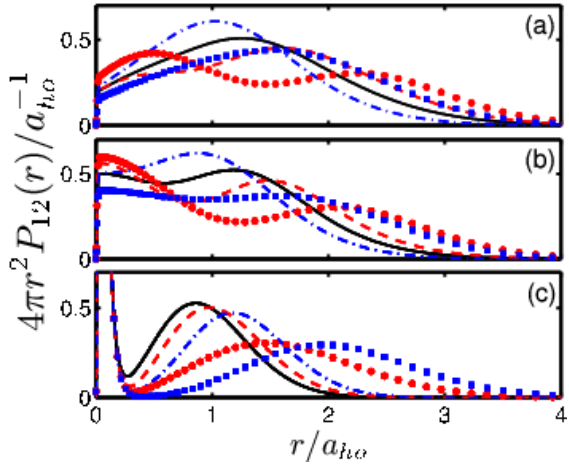}
\caption{(Color Online) Pair distribution function $P_{12}(r)$,
multiplied by $r^2$, for
two-component Fermi gases
with $\kappa=4$ for different scattering lengths $a_s$:
(a) $a_s=-a_{ho}$ (BCS regime), (b) $1/a_s=0$ (unitarity),
and (c) $a_s=0.1a_{ho}$ (BEC regime).
Dashed and dash-dotted lines
show $P_{12}(r)r^2$ for $N=3$ (two heavy particles) with
$L=0$
and $1$, respectively.
Circles and squares show $P_{12}(r)r^2$ for $N=3$ (two light particles) with
$L=0$ and $1$, respectively.
 Solid lines show $P_{12}(r)r^2$ for $N=4$ with $L=0$.
}\label{Cor12k4}
\end{figure}

Next, we analyze how the behaviors of the pair distribution
functions $P_{12}(r)$ for $N=3$ and $4$ change along the crossover
if the mass ratio is changed from $\kappa=1$ to $4$.
Figure~\ref{Cor12k4} shows the pair distribution functions for
$\kappa=4$. For $N=3$, we consider three-particle systems with
either a spare light particle or with a spare heavy particle. The
pair distributions for the three-particle system with two light
particles and one heavy particle are notably broader than those for
the three-particle system with one light particle and two heavy
particles. This behavior can be attributed to the fact that
$a_{ho}^{(1)} > a_{ho}^{(2)}$. Besides this, a comparison of the
pair distribution functions shown in Fig.~\ref{Cor12k4} for
$\kappa=4$ and those shown in Fig.~\ref{Corr3and4} for $\kappa=1$
reveals that the overall behavior of the $P_{12}(r)$ is similar.

\begin{figure}[h]
\includegraphics[scale=0.6]{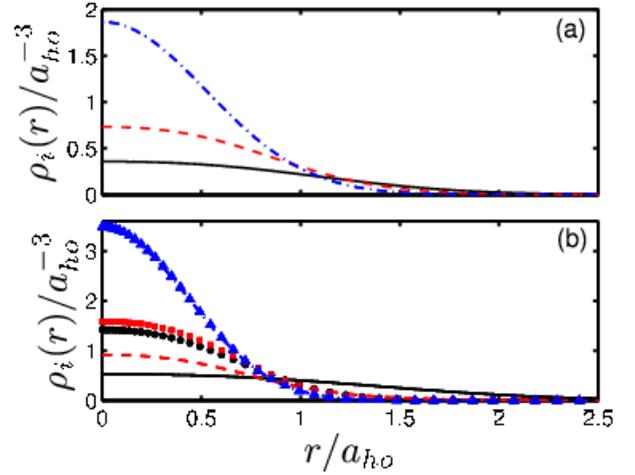}
\caption{(Color Online) One-body densities $\rho_1(r)$ and
$\rho_2(r)$ for $N=4$ and (a) $\kappa=1$ and (b) $\kappa=4$ for
different scattering lengths $a_s$ [for $\kappa=1$, $\rho_1(r)$ and
$\rho_2(r)$ coincide and only $\rho_2(r)$ is shown]: Circles and
solid lines show $\rho_1(r)$  and $\rho_2(r)$ for $a_s=0$, squares
and dashed lines show $\rho_1(r)$  and $\rho_2(r)$ for $1/a_s=0$,
and triangles and dash-dotted lines show $\rho_1(r)$ and $\rho_2(r)$
for $a_s=0.1a_{ho}$. Note, $\rho_2(r)$ for $\kappa=4$ and $a_s=0$
[solid line in panel~(b)] is multiplied by a factor of three to
enhance the visibility.}\label{den4}
\end{figure}

Figures~\ref{den4}(a) and (b) show the one-body densities for
$\kappa=1$ and $4$, respectively. In the non-interacting limit [the
solid lines show $\rho_1(r)$ and the circles show $\rho_2(r)$], the
sizes of $\rho_1(r)$ and $\rho_2(r)$ are determined by
$a_{ho}^{(1)}$ and $a_{ho}^{(2)}$, respectively. As is evident in
Fig.~\ref{den4}, the density of the light particles extends to
larger $r$ than the density of the heavy particles. The density
mismatch for $\kappa=4$ between the two one-body densities decreases
as $a_s$ is tuned through the strongly-interacting regime to the
weakly-interacting molecular BEC side. In the weakly-interacting
molecular BEC regime, two molecules consisting each of a heavy and a
light particle form. In this regime, the size of the system is
determined by the molecular trap length and the densities
$\rho_{1}(r)$ and $\rho_{2}(r)$ [triangles and dash-dotted line in
Fig.~\ref{den4}(b)] nearly coincide. Furthermore, the densities are
to a very good approximation described by the one-body density for
two bosonic molecules of mass $m_1+m_2$ interacting through an
effective repulsive interaction characterized by the dimer-dimer
scattering length ($a_{dd}\approx 0.77a_s$ for
$\kappa=4$~\cite{petr05,vonstechtbp}).

We have also analyzed the
pair distribution functions $P_{ii}(r)$ for $\kappa=1$ and $4$ (not
shown). The small $r$ region of the $P_{ii}(r)$ is controlled by the
Pauli exclusion principle between identical fermions. In the
weakly-interacting molecular
BEC regime, the pair distribution functions $P_{11}(r)$ and
$P_{22}(r)$ nearly coincide even for $\kappa=4$. In this regime, the
pair distribution functions $P_{ii}(r)$ are well approximated by
that for two particles of mass $m_1+m_2$ interacting with a
repulsive potential characterized by $a_{dd}$.

\subsection{Structural properties at unitarity}
\label{sec_struc2}

This section discusses selected structural properties of
two-component equal mass Fermi gases at unitarity with up to $N=20$
atoms. For small systems ($N \le 6$), we present structural
properties calculated using both the CG and the FN-DMC methods. For
larger systems, however, our interpretation relies solely on the
structural properties calculated by the FN-DMC method.

\begin{figure}[h]
\includegraphics[scale=0.6]{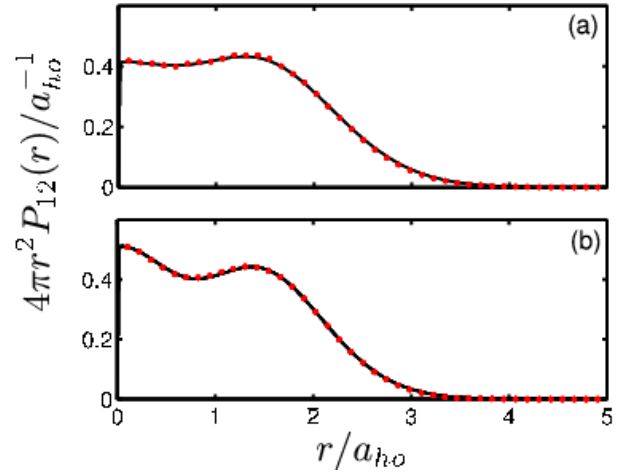}
\caption{(Color Online) Pair distribution functions $P_{12}(r)$,
multiplied by $r^2$, at unitarity for equal mass Fermi systems with
(a) $N=3$ ($L=1$) and (b) $N=4$ ($L=0$) atoms calculated by the CG
method (solid lines) and by the FN-DMC method (circles). The
agreement is excellent. }\label{CorrComp}
\end{figure}

To assess the accuracy of the nodal surfaces employed in our FN-DMC
calculations as well as of the accuracy of the mixed estimator [see
Eq.~(\ref{eq_mixed}) in Sec.~\ref{sec_fndmc}],
Figs.~\ref{CorrComp}(a) and (b) compare the pair distribution
functions $P_{12}(r)$ for the three-particle system with $L=1$ and
the four-particle system with $L=0$, respectively, calculated by the
CG and the FN-DMC methods.
The agreement between the pair distribution functions calculated by
the CG method (solid lines) and the FN-DMC method (circles) is very
good, validating the construction of the nodal surface of
$\psi_{T}$. Furthermore, the good agreement suggests that the mixed
estimator results, for the guiding functions employed, in structural
properties very close to those one would obtain by an exact
estimator.

\begin{figure}[h]
\includegraphics[scale=0.6,angle=-0]{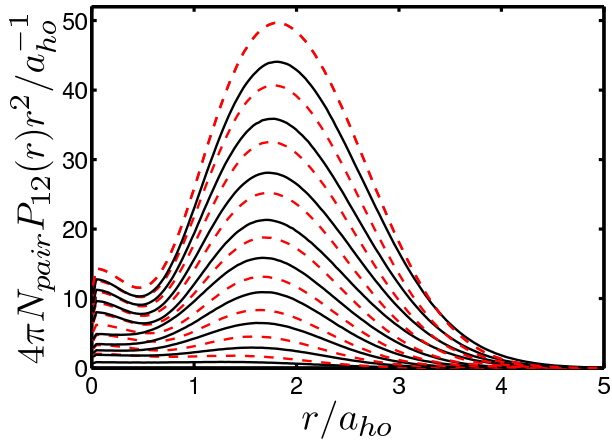}
\caption{(Color Online) Dashed and solid lines show the pair
distribution functions $P_{12}(r)$, multiplied by $r^2 N_{pair}$
($N_{pair}$ denotes the number of interspecies distances), for equal
mass Fermi systems at unitarity with even $N$ ($N=4,6,\cdots,20$)
and odd $N$ ($N=3,5,\cdots,19$), respectively, calculated by the
FN-DMC. Beyond $r \approx a_{ho}$, $P_{12}(r) r^2 N_{pair}$ is
smallest for $N=3$ and largest for $N=20$. }\label{pair_largen}
\end{figure}

Figure~\ref{pair_largen} shows the pair distribution functions
$P_{12}(r)$ calculated by the FN-DMC method for equal mass Fermi
systems with $N=3-20$ at unitarity.
To simplify the comparison, Fig.~\ref{pair_largen} shows the even
$N$ results as a dashed line and the odd $N$ results as a solid
line. Furthermore, $P_{12}(r) r^2$ is multiplied by the number
$N_{pair}$ of interspecies distances so that the $N=3$ distribution
function has the smallest and the $N=20$ distribution function the
largest amplitude for $r \gtrsim a_{ho}$. The pair distribution
functions for even $N$ show a similar behavior for all $N$
considered; both the small $r$ and the large $r$ peaks grow
monotonically and smoothly with increasing $N$. For odd $N$, in
contrast, the small $r$ peak changes somewhat discontinuously at $N
\approx 11$. This behavior can be attributed to the guiding
functions employed. For even $N$, the guiding function $\psi_{T1}$,
whose nodal surface is constructed from the two-body solution, gives
the lowest energy for all $N$ (except for $N=4$). For odd $N$,
however, $\psi_{T2}$ results in a lower energy for $N\le 9$,
$\psi_{T3}$ for $N=11$, and $\psi_{T1}$ for $N \ge 13$. Thus, the
pair distribution functions clearly reveal how the structural
properties depend on the nodal surface employed in the FN-DMC
calculations and provide much deeper insights into the different
$\psi_T$ employed than a mere comparison of the energies.

For $N \ge 13$, Fig.~\ref{pair_largen} indicates that the amplitudes
of the scaled interspecies pair distribution functions are nearly
the same for neighboring systems. For example, the quantities
$P_{12}(r)r^2 N_{pair}$ for $N=18$ and 19 agree to a good
approximation, suggesting that one can think of the $N=18$ system as
consisting of nine pairs, and of the $N=19$ system as consisting of
nine pairs plus a spare atom. Note that this interpretation hinges
critically on the nodal surface employed in our FN-DMC calculations;
a small change in the nodal structure of the guiding function may
change the small $r$ behavior of the pair distribution functions
non-negligibly.

\begin{figure}[h]
\includegraphics[scale=0.6,angle=0]{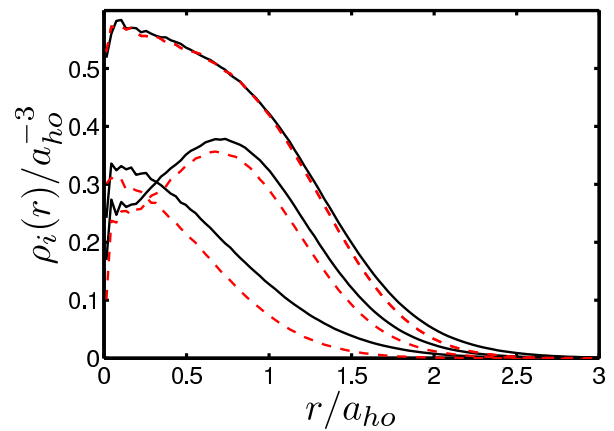}
\caption{(Color Online)
The one-body density $\rho_1(r)$ (solid
lines) is shown for $N=3$, 9 and 15
(for $r > 0.5 a_{ho}$, from bottom to
top), together with the
one-body density $\rho_2(r)$ (dashed
line) for $N=3$, 9 and 15 (for $r > 0.5 a_{ho}$, from bottom to top)
for equal mass two-component Fermi gases at unitarity calculated by
the FN-DMC method. }\label{onebody}
\end{figure}

We next investigate in Fig.~\ref{onebody}
where the spare particle is located
in the odd-$N$ systems at
unitarity.
This figure shows the one-body densities $\rho_1(r)$ (solid lines)
and $\rho_2(r)$ (dashed lines) for $N=3$, 9 and 15. For $N=3$, the
difference between $\rho_1(r)$ and $\rho_2(r)$ is roughly constant
across the trap. The behavior is similar for $N=9$. Interestingly,
the densities for $N=9$ show a minimum at $r=0$, reflecting the fact
that the FN-DMC calculations employ the nodal surface of the ideal
Fermi gas, i.e., use $\psi_{T2}$ [Eq.~(\ref{eq_t2})]. For $N=15$,
the nodal surface employed is constructed from the two-body solution
[see Eq.~(\ref{eq_t1})], and consequently, the behavior of the
density profiles differs from that for the smaller $N$. The
densities $\rho_1(r)$ and $\rho_2(r)$ nearly coincide at small $r$.
At large $r$, however, the density $\rho_1(r)$ has a larger
amplitude than $\rho_2(r)$ (recall $N_1=N_2+1$). Our data for $N=15$
indicate that the spare particle is not distributed uniformly
throughout the trap but has an increased probability to be found
near the edge of the cloud. Possible consequences of this finding
for the excitation gap have already been discussed in
Refs.~\cite{blumePRL07,son07}.

\begin{figure}[h]
\includegraphics[scale=0.6,angle=0]{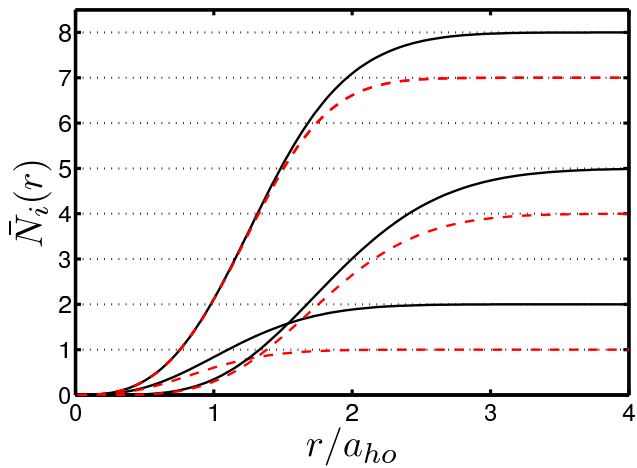}
\caption{(Color Online)
Solid and dashed lines show the integrated quantities
$\bar{N}_1(r)$ and $\bar{N}_2(r)$ [Eq.~(\ref{eq_nbar})], respectively,
as a function of $r$
for a two-component Fermi gas at unitarity.
At large $r$,
the curves correspond from bottom to top
to $N=3$, 9 and 15.
}
\label{integrate}
\end{figure}

To quantify the analysis of the one-body densities,
we integrate $\rho_i(r)$ over $r$,
\begin{eqnarray}
\label{eq_nbar} \bar{N}_i(r)=4 \pi \int_0^r \rho_i(r') r'^2 dr'.
\end{eqnarray}
For a finite upper integration limit, $\bar{N}_i(r)$ monitors how
many of the $N_i$ particles are located between zero and $r$.
Figure~\ref{integrate} shows $\bar{N}_i(r)$ for $N=3$, 9 and 15. As
in Fig.~\ref{onebody}, the results for component one are shown by
solid lines and those for component two by dashed lines; in the
large $r$ limit, the $\bar{N}_i(r)$ equal $N_i$, as expected. One
can now read off nicely,
in which $r$-regions the densities of the two components agree and
where they disagree. For $N=3$, e.g., the two atoms of component one
and the one atom of component two are added over approximately the
same $r$-region. For $N=15$, in contrast, the first five atoms of
the two components are located in the region with $r\lesssim
1.5a_{ho}$; this core region can be considered ``fully paired''. The
last three atoms of component one and the last two atoms of
component two form, loosely speaking, a ``partially paired or
unpaired outer shell''. We find similar behaviors for the odd-$N$
systems with $N=13$, 17 and 19. It will be interesting to see if
this interpretation holds for larger $N$, and if this information
can be used to shed light on the phase diagram of asymmetric Fermi
gases~\cite{zwie06,part06}.

\begin{figure}[h]
\includegraphics[scale=0.5]{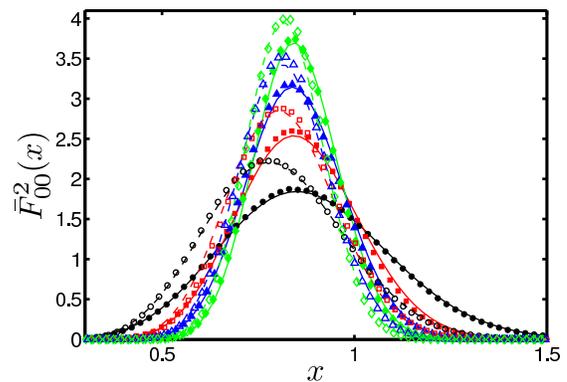}
\caption{(Color Online) The hyperradial density
$\bar{F}_{00}^2(x)$ is shown as a
function of the dimensionless hyperradius $x$ for $N=3-10$,
calculated using the mixed Monte Carlo estimator (symbols) and the
analytical expression with the FN-DMC energies (lines),
respectively. The maximum of $\bar{F}_{00}^2(x)$ is smallest for
$N=3$ and largest for $N=10$; the MC results are shown by filled
circles for $N=3$, open circles for $N=4$, filled squares for $N=5$,
open squares for $N=6$, filled triangles for $N=7$, open triangles
for $N=8$, filled diamonds for $N=9$ and open diamonds for $N=10$.}
\label{fig_hyper_mc}
\end{figure}

To further verify the validity of the special properties of
two-component Fermi gases at unitarity as well as to further assess
the accuracy of our guiding functions employed in the FN-DMC
calculations, we analyze the hyperradial densities $\bar{F}_{00}^2(x)$
for various $N$.
Symbols in Fig.~\ref{fig_hyper_mc} show the
dimensionless hyperradial density $\bar{F}_{00}^2(x)$ calculated
using the mixed Monte Carlo estimator, Eq.~(\ref{eq_mixed}),
for $N=3$ to 10. Here, $x$ is the
dimensionless
hyperradius defined just above Eq.~(\ref{Hx}) and the normalization
of the $\bar{F}_{00}$ is chosen so that $\int_0^{\infty}
\bar{F}_{00}^2(x) dx=1$. The dimensionless hyperradius $x$ is scaled
by $R'_{NI}$, which we evaluate by approximating $E_{NI}$ in
Eq.~(\ref{eq_rnihyper}) by $E_{NI,EFT}$. This is similar to the
``smoothing procedure'' discussed in the context of Figs.~\ref{gapN}
and \ref{CN}. The hyperradial densities become more compact as $N$
increases, owing to the increase of the effective mass $\mu_{eff}$
entering into the effective hyperradial Schr\"odinger equation
[Eq.~(\ref{Hx})] with increasing $N$. Furthermore, the maximum of
the hyperradial densities occurs at slightly larger $x$ values for
odd $N$ systems than for even $N$ systems, in agreement with the
odd-even staggering discussed in Sec.~\ref{sec_unitary} in the
context of the hyperradial potential curves $V(R)$.

In the limit of zero-range interactions, the adiabatic approximation
is expected to be exact (see Sec.~\ref{sec_hyper}). In this case,
the functional form of the hyperradial wave functions is known
analytically [see Eq.~(\ref{eq_fhyper})], and can be compared with
the Monte Carlo results obtained for short-range potentials by
sampling the total wave function and integrating over all
coordinates but the hyperradius. Solid lines in
Fig.~\ref{fig_hyper_mc} show the hyperradial densities
$\bar{F}_{00}^2(x)$ for $N=3$ to $10$ predicted analytically for
zero-range interactions, using the FN-DMC energies $E_{00}$ listed
in Table~\ref{tab1} of this paper and Table~II of
Ref.~\cite{vonstechtbp}. The agreement between the analytical
results and the Monte Carlo results obtained for finite range
potentials is quite good. On the one hand, this agreement lends
numerical support for the separability or near separability of the
total wave function into a hyperradial and a hyperangular part. On
the other hand, the good agreement suggests that the nodal surface
employed in our MC calculations is appropriate.

\begin{figure}[h]
\includegraphics[scale=0.6]{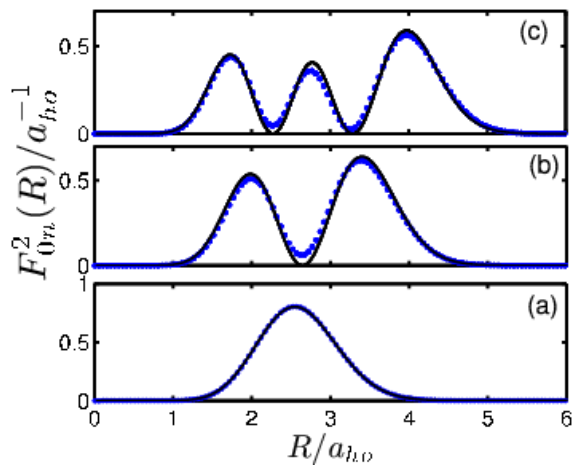}
\caption{(Color Online) Hyperradial density $F_{0n}^2(R)$ for
$n=1,2$ and $3$. Here, we choose $\mu_N=m$ ($\mathcal{L}=a_{ho}$).
The solid lines show the analytical solutions while the circles show
the numerical results obtained by integrating $(\Psi^{rel})^2$
calculated by the CG method over all coordinates but the hyperradius
$R$.}\label{HRWF6p}
\end{figure}

Finally, we analyze the hyperradial densities calculated by the CG
approach for the $N=6$ system. Since the CG approach allows for the
determination of excited states, this analysis allows us to verify
that the $2\hbar\omega$ spacing reported in Ref.~\cite{blumePRL07}
and discussed in Sec.~\ref{sec_hyper} corresponds indeed to
breathing-mode excitations, i.e., to excitations along the
hyperradial coordinate. To extract the hyperradial densities, we
integrate the square of the wavefunction $\Psi^{rel}$ over all the
coordinates but the hyperradius. If the universal behavior is
fulfilled, then the hyperradial densities should coincide with the
square of the analytically determined $F_{\nu n}(R)$,
Eq.~(\ref{eq_fhyper}), which are shown in Fig.~\ref{HRWF6p} by solid
lines. The integration over the hyperangular coordinates is carried
out using Monte Carlo integration techniques. Symbols in
Fig.~\ref{HRWF6p} show the resulting hyperradial densities $F_{\nu
n}^2(R)$ for $\nu=0$ and $n=0, 1$ and $2$. The numerically
determined hyperradial densities indicate that the excitations are
to a good approximation located along the $R$ coordinate, supporting
the interpretation of the $2 \hbar \omega$ spacing within the
hyperspherical framework. The agreement between the numerical and
analytical results is excellent for the ground state [see
Fig.~\ref{HRWF6p}(a)]. For the excited states with $n=1$ and $2$,
the small deviations between the numerical and analytical results
may be due to finite range effects or not fully converged numerical
results.

\section{Conclusions}
\label{sec_conclusion}

This paper presents a microscopic picture of the properties of
ultracold two-component fermionic systems in a trap. Complementing
previous studies~\cite{vonstechtbp,blumePRL07}, we focus on the
energetics of odd $N$ systems, and the structural properties of both
odd and even $N$ systems.

For sufficiently few particles, we solve the Schr\"odinger equation
for equal and unequal mass systems,
starting from a model Hamiltonian with
short-range interspecies $s$-wave interactions using the CG approach.
This basis set expansion technique allows for the determination of
the eigenspectrum and eigenstates with controlled accuracy
throughout the BEC-BCS crossover. We find that the spectrum and the
structural properties of small trapped two-component Fermi systems
change qualitatively throughout the crossover regime.

An analysis of the energies of the $N=3$ systems in the
weakly-interacting BEC and BCS regimes allows us to determine the
validity regime of the analytically determined perturbative
expressions for small $|a_s|$. Furthermore, we find that the angular
momentum of the $N=3$ ground state changes from $L=1$ in the
weakly-attractive BCS regime to $L=0$ in the weakly-repulsive BEC
regime for all mass ratios considered. By additionally treating the
$N=2$ and 4 systems, we determine the excitation gap $\Delta(3)$
throughout the crossover region: For equal frequencies, the
excitation gap decreases for all scattering lengths with increasing
mass ratio. For $N=4$ systems with $\kappa \le 13$, we determine the
$L=0$ excitation spectrum at unitarity. The spectrum determines the
$s_{\nu}$ coefficients of the hyperradial potential curves and also
verifies within our numerical accuracy the $2 \hbar \omega$ spacing
prediction, which was derived analytically assuming
universality~\cite{wern06}. We verified in a number of cases that
the $2 \hbar \omega$ spacing corresponds indeed to breathing-mode
excitations.

Our analysis of the energetics is complemented by studies of the
structural properties. For the four-particle system with equal and
unequal masses, e.g., we show how the pair distribution functions in
the $a_s>0$ region (small $a_s$) can be described by a system of two
molecules interacting through an effective dimer-dimer potential
with positive dimer-dimer scattering length. A similar analysis was
carried out for the $N=3$ system and we verified that this system
behaves as an interacting system of an atom and a dimer.

Our small $N$ studies have implications for optical lattice
experiments. Our results can be applied directly if each optical
lattice site is approximately harmonic in the non-tunneling regime.
In this context, Ref.~\cite{kestnerPra07} already pointed out,
including the energies of the two- and three-particle system, that
the occupation of optical lattice sites with three equal-mass atoms
is unfavorable. To start with let us consider a system with two
lattice sites and six atoms (three of each species). We imagine that
the lattice sites are loaded by adiabatically turning up the barrier
height between the two sites. It follows from our energies
calculated for $N=2$ through $4$, that the system ends up with
unequally occupied lattice sites at the end of the ramp: For both
equal and unequal masses, and for all scattering lengths, the energy
of the two-site lattice is minimal if two atoms occupy one site and
the other four atoms occupy the other site. This is simply a
consequence of the fact that $\Delta(3)$ is positive throughout the
crossover (note, the energy of the unequal mass system might be
further lowered if we consider the formation of pentamers and
sextamers with negative energies; these states are not included in
our analysis). The arguments presented here for just two lattice
sites generalize readily to lattices with multiple sites.

Instead of ramping up the barrier height adiabatically, we now
imagine a fast non-adiabatic ramp. In this case, the likelihood of
finding three atoms per lattice site at the end of the ramp is
finite. Since the excitation frequencies for two-, three- and
four-particle systems are different, a ``purification
sweep''~\cite{thal06} can be used to then prepare a system with
either three or no particles per site. These three-particle systems
could be investigated spectroscopically (see, e.g.,
Sec.~\ref{sec_exc} for a discussion of the excitation spectrum of
the four-particle system). Alternatively, one might ask whether it
would be possible to measure the odd-even physics by adiabatically
lowering the lattice barrier and monitoring the point at which tunneling
sets in.

In addition to systems with equal numbers of atoms in the two
species, we consider an optical lattice with twice as many heavy as
light atoms. If the mass ratio is sufficiently large, trimers
consisting of two heavy atoms and one light atom with negative
energy can form at each lattice site, paving the way for
spectroscopic studies of these delicate systems. Furthermore, by
starting with a bound trimer in a deep lattice and then lowering the
lattice height, a gas consisting of bound trimers can possibly be
prepared.

To extend the studies of the energetics and structural properties to
larger systems, we employed the FN-DMC technique. This approach determines
the lowest energy of a state that has the same symmetry as a so-called
guiding function and thus an upper bound to the exact eigenenergy.
Detailed comparisons of the
energies and the structural properties calculated by the CG
and FN-DMC approaches benchmark the nodal surfaces employed for
systems with $N \le 6$. In the strongly-correlated
unitary regime, e.g., the FN-DMC energies for equal-mass two-component Fermi
systems
agree with the CG energies to within 2\%.

Our even $N$ energies ($N \le 30$) for equal-mass systems at
unitarity show vanishingly small shell structure. Applying the local
density approximation and approximating the non-interacting energies
by the corresponding extended Thomas-Fermi expression, we find
$\xi_{tr}=0.467$, which is somewhat larger than the value of the
homogeneous system, $\xi_{hom}=0.42$. We note that the expression
$\sqrt{\xi_{tr}}E_{NI,EFT}$ describes the equal-mass energies for
$N\le30$ at unitary very well; the small disagreemet between
$\xi_{tr}$ and $\xi_{hom}$ is most likely due to the small number of
particles considered in the present work. Combining the even and odd
$N$ energies, we find that the excitation gap $\Delta(N)$ at
unitarity increases with $N$. Also, the one-body densities and pair
distribution functions at unitarity are studied for up to $N=20$.
For odd $N$ with $N \gtrsim 11$, we observe that the extra
``unpaired'' particle is located predominantly near the edge of the
cloud, in agreement with previous
predictions~\cite{blumePRL07,son07}. This suggests that the LDA
cannot be applied to determine the excitation gap. Furthermore, we
find that the hyperradial densities of the lowest gas-like state
with $N \le 10$ calculated for short-range interactions by the
FN-DMC method agree with the analytically predicted ones, indicating
that the lowest gas-like state does indeed behave
universally. Selected
hyperradial densities for larger $N$ were already presented in
Ref.~\cite{blumePRL07}.

The energies and structural properties for the equal-mass
two-component Fermi systems at unitarity presented in this paper may
serve as a benchmark for other calculations. Recently, e.g., a DFT
treatment determined the energies for systems with up to $N=20$
particles~\cite{bulgac07}. The good agreement between the FN-DMC
energies and the DFT energies suggests that the functional employed
in the DFT calculations captures the key physics. However, close
inspection of the FN-DMC and DFT energies indicates that the
agreement between the even $N$ and odd $N$ energies is not equally
good. While this could be a consequence of the nodal surfaces
employed in our FN-DMC calculations, it could alternatively indicate
that the DFT treatment employed in Ref.~\cite{bulgac07} for odd $N$
is not optimal. Thus, it is hoped that our results will proof
helpful in assessing the accuracy of the DFT approach and other
approaches.

We acknowledge support by the NSF, and fruitful discussions
with J. D'Incao and S. Giorgini.


\end{document}